\documentclass[fleqn,usenatbib]{mnras}

\usepackage{newtxtext,newtxmath}

\usepackage[T1]{fontenc}
\usepackage{hyperref}
\usepackage{breakurl}
\usepackage{ulem}
\DeclareRobustCommand{\VAN}[3]{#2}
\let\VANthebibliography\thebibliography
\def\thebibliography{\DeclareRobustCommand{\VAN}[3]{##3}\VANthebibliography}

\usepackage{graphicx}	

\title[Abundant sub-micron grains in EDDs]{Abundant sub-micron grains revealed in newly discovered extreme 
debris discs}

\author[A. Mo\'or et al.]{
Attila Mo\'or,$^{1,2}$\thanks{E-mail: moor.attila@csfk.org}
P\'eter \'Abrah\'am,$^{1,2,3,4}$
Kate Y.~L. Su,$^{5}$
Thomas Henning,$^{6}$
Sebastian Marino,$^{7}$
Lei Chen,$^{1,2}$ 
\newauthor
\'Agnes K\'osp\'al,$^{1,2,6,3}$
Nicole Pawellek,$^{4,1,2}$
J\'ozsef Varga$^{1,2}$
and Kriszti\'an Vida$^{1,2}$
\\
$^{1}$Konkoly Observatory, Research Centre for Astronomy and Earth Sciences, HUN-REN,\\ 
Konkoly-Thege Mikl\'os \'ut 15-17, H-1121 Budapest, Hungary \\
$^{2}$CSFK, MTA Centre of Excellence, Budapest, Konkoly Thege Mikl\'os \'ut 15-17., H-1121, Hungary\\
$^{3}$ELTE E\"otv\"os Lor\'and University, Institute of Physics, P\'azm\'any P\'eter s\'et\'any 1/A, H-1117 Budapest, Hungary\\
$^{4}$Department of Astrophysics, University of Vienna, T\"urkenschanzstra{\ss}e 17, 1180, Vienna, Austria\\
$^{5}$Department of Astronomy/Steward Observatory, The University of  Arizona, Tucson, AZ 85721-0009, USA\\
$^{6}$Max-Planck-Institut f\"ur Astronomie, K\"onigstuhl 17, D-69117 Heidelberg, Germany\\
$^{7}$Department of Physics and Astronomy, University of Exeter, Stocker Road, Exeter EX4 4QL, UK \\
}

\date{Accepted XXX. Received YYY; in original form ZZZ}

\pubyear{2024}

\begin{document}
\label{firstpage}
\pagerange{\pageref{firstpage}--\pageref{lastpage}}
\maketitle

\begin{abstract}
Extreme debris discs (EDDs) are bright and warm circumstellar dusty structures around main sequence stars.
They may represent the outcome of giant collisions occuring in the terrestrial region between large planetesimals 
or planetary bodies, and thus provide a rare opportunity to peer into the aftermaths of these events. Here, we 
report on results of a mini-survey we conducted with the aim to increase the number of known EDDs, investigate 
the presence of solid-state features around 10\,{\micron} in eight EDDs, and classify them into the silica or silicate 
dominated groups. We identify four new EDDs and derive their fundamental properties. For these, and for four other 
previously known discs, we study the spectral energy distribution around 10\,{\micron} by means of VLT/VISIR photometry 
in three narrow-band filters and conclude that all eight objects likely exhibit solid-state emission features from sub-micron 
grains. We find that four discs probably belong to the silicate dominated subgroup. Considering the age distribution of 
the entire EDD sample, we find that their incidence begins to decrease only after 300\,Myr, suggesting that the earlier 
common picture that these objects are related to the formation of rocky planets may not be exclusive, and that other 
processes may be involved for older objects ($\gtrsim$100\,Myr). Because most of the older EDD systems have wide, eccentric 
companions, we suggest that binarity may play a role in triggering late giant collisions.

\end{abstract}




\section{Introduction}  \label{sec:intro}

In recent decades, observations at mid-infrared (mid-IR) wavelengths have led to the identification of a 
number of main-sequence stars surrounded by warm dust at temperatures of $\sim$200--600\,K \citep[e.g.,][]{kennedy2013,fernando2014,cotten2016}. 
For Sun-like stars 
with a spectral type between mid F and late K, these temperatures are indicative of dust grains located within 
a few au, i.e. in a region where the terrestrial planets orbit in the Solar System.     
The lifetime of such small warm particles is significantly shorter than the age of their host star, 
as under 
the influence of stellar radiation and stellar wind they are removed from the system on a timescale of tens of 
thousands of years at most.
So the long-term maintenance of a circumstellar disc requires continuous dust replenishment from erosion 
of larger bodies. The second generation debris grains could either be released from an in situ collisional cascade 
that grinds a ring of planetesimals into (sub)micron-sized particles, from a single large collision, 
or even from bodies originally located much further away, for example as sublimation or 
disruption of icy minor bodies transported from a cold, outer reservoir into the inner region 
\citep{wyatt2008,rigley2022}.

Some warm debris discs contain such large amounts of dust that are clearly not sustainable for the lifetime of the system, 
implying instead that a recent episodic dust production event has resulted in a strongly elevated dust level for a 
temporary period \citep{wyatt2007}. The IR luminosity of the dustiest of these warm transient discs exceeds 1\% of the 
luminosity of their host stars. These so-called extreme debris discs (EDDs), of which we currently know about 
20 \citep{oudmaijer1992,gorlova2004,song2005,gorlova2007,rhee2008,melis2010,zuckerman2012,dewit2013,tajiri2020,melis2021,moor2021,higashio2022}, 
are proposed to be signatures of giant impacts during the final assembly of rocky planets, 
or the outcome of a late dynamical instability of a planetary system, or 
of a collision between a rocky planet and its former moon that became unbound due to tidal evolution
\citep{melis2016,su2019,moor2021,melis2021,hansen2023}.

Mid-IR spectroscopic observations between 5 and 35\,{\micron} with the Infrared Spectrograph \citep[IRS;][]{houck2004} onboard 
the {\sl Spitzer} Space Telescope \citep[{\sl Spitzer};][]{werner2004} revealed several warm debris discs with solid state emission features that can mostly be attributed to small silicate particles with sizes from submicron to a few {\micron} \citep[e.g.,][]{ballering2014,mittal2015}.  
Position and shape of the observed features allow us to study the crystalline fraction and mineralogy of dust which can provide 
insights into the composition of the parent bodies (the outer layers for larger objects) and can help to constrain the processes 
that led to the dust release \citep{henning2010,lisse2008,lisse2012,olofsson2012}. Spectra with the most prominent features tend to be 
displayed by bright transient debris discs. 
By analysing the IRS data of 12 debris discs, most having transient nature, \citet{morlok2014} found that 
basically two groups can be distinguished. For a subset of discs, the most dominant feature peaks between 9 and 9.5\,{\micron} 
indicating the presence 
of silica grains \citep[e.g. quartz, cristobalite, silica glass,][]{koike2013}, while for the other subset, the spectra are charaterized by strong features peaking in the 9.9--11.3\,{\micron} 
wavelength range, suggesting dust material dominated by silicate particles \citep[e.g. pyroxene, olivine,][]{henning2010}. By comparing these spectra with those of 
terrestrial and Martian rocks, 
they argued that the former subset is more similar to crustal material, while the latter has features more akin to
mantle-type samples. Assuming that the source of the observed debris dust was a collision between differentiated planetary bodies, 
the above comparison gives an indication of the portion of the planet from which the ejected material originated and can also provide 
a constraint on the nature of the collision \citep{morlok2014}.

EDDs offer a direct view at the aftermaths of large collisions. Here we present a study, whose aim was to 
gain a deeper understanding of this phenomenon. First we focus on increasing the available EDD sample, and report on 
the discovery of four new EDDs. Then we supplement the newly discovered sources with four already known 
objects, and perform a mini-survey using the VLT/VISIR camera to obtain multi-band mid-IR 
photometry for eight EDDs, with the aim to investigate whether they also show solid-state features around 
10\,{\micron}, and which of the above subgroups they are more likely to belong to. The novelty of this project 
is that before 2009, when the primary "cold" mission of the {\sl Spitzer} stopped, and thus the IRS ceased to operate, 
only a handful of EDDs were known. Although some of the brightest objects identified since 2009 have been 
observed by means of ground-based mid-IR spectroscopy in the 8--13\,{\micron} range, only half of the currently 
known EDDs have mid-IR spectroscopic data. Our survey provides narrow-band photometry for sources which would otherwise 
be too faint for ground-based spectroscopy, this way significantly increasing the sample of EDDs with information on the 
10\,{\micron} feature.

Basic properties of the newly discovered EDDs are derived and presented in Sect.~\ref{sec:newlydiscovered}.
The mid-IR spectroscopic observations and data reduction are 
summarized in Sect.~\ref{sec:visirobs}. The results are presented in Sect.~\ref{sec:results}
and their implication are discussed in Sect.~\ref{sec:discussion}. Finally, in Sect.~\ref{sec:summary}, the main
outcomes of our investigation are summarized.

\section{Properties of the newly discovered EDDs} \label{sec:newlydiscovered}
We have been carrying out a long term project whose aim is to identify and investigate EDDs around 
Sun-like stars located within 400\,pc using a combined data set, based on the AllWISE mid-IR
photometric \citep{wright2010} and Gaia Data Release 3 \citep[Gaia~DR3,][]{gaia2023} astrometric catalogues.
All details of the identification process will be described in a subsequent paper (Mo\'or et al. in prep.). 
For our photometric survey, we selected four, yet unpublished objects from our comprehensive 
EDD list, J060917.00--150808.5 (hereafter J060917), J071206.54--475242.3 (J071206), J104416.70--
451613.9 (J104416), and J204315.23$+$104335.3 (J204315), on the basis of their mid-IR 
brightness and observability from the southern hemisphere.

The four newly discovered EDDs are complemented by four recently identified objects from the literature.
TYC 5940-1510-1 (TYC\,5940), TYC 8105-370-1 (TYC\,8105), and TYC 4946-1106-1 (TYC\,4946) are those three 
of the six new EDDs presented in \citet{moor2021} that are observable from
Cerro Paranal, while J092521.90-673224.8 (J092521) is taken from
\citet{higashio2022}. 

In the following, we derive fundamental stellar and disc properties of the four 
newly discovered EDDs and J092521. Similar data for TYC\,5940, TYC\,8105, and TYC\,4946 can
be found in \citet{moor2021}.

\subsection{Fundamental stellar properties} \label{sec:stellarprops}
To estimate fundamental stellar properties of the four newly discovered EDDs and J092521 we obtained their high resolution 
($R=48\,000$) optical spectra using the Fibre-fed Extended Range Optical Spectrograph \citep[FEROS,][]{kaufer1999}
mounted on ESO/MPG 2.2\,m telescope at La Silla Observatory (Chile). 
The observations were executed in three different projects (090.C-0815(A) -- PI: Attila Mo\'or, 0106.A-9012(A) -- PI: Sebastian Marino, and 0109.A-9029(A) -- PI: Thomas Henning). In all measurements we used the object-sky observing mode in which one of the two fibres targeted the star while the other was positioned at the sky. The log of observations is given in Table~\ref{tab:feroslog}.
The data reduction of the obtained spectra was carried out utilizing the CERES (Collection of Elemental Routines for 
Echelle Spectra) pipeline developed by \citet{brahm2017a}. 

\begin{table}                                                                  
\begin{center} 
\caption{Log of spectroscopic observations. \label{tab:feroslog}}
\begin{tabular}{lcc}                                                     
\hline\hline
AllWISE / Gaia DR3 name & Obs. date & Programme ID \\
\hline
J060917.00$-$150808.5 & 2021-02-13 & 0106.A-9012(A) \\
J071206.54$-$475242.3 & 2021-02-14 & 0106.A-9012(A)\\
J092521.90$-$673224.8 & 2012-12-25 & 090.C-0815(A)\\
        & 2022-06-08 & 0109.A-9029(A)\\
J104416.70$-$451613.9 & 2021-02-13 & 0106.A-9012(A)\\
Gaia~DR3~5366814012930376960 & 2021-02-13 & 0106.A-9012(A) \\
J204315.23$+$104335.3 & 2022-06-09 & 0109.A-9029(A)\\
\hline
\\
\end{tabular}
\end{center}
\end{table}

We employed the ZASPE tool \citep[Zonal Atmospheric Stellar Parameters Estimator,][]{brahm2017b} to derive 
stellar atmospheric parameters, the effective temperature ($T_{\rm eff}$), metallicity ([Fe/H]),
surface gravity ($\log{g}$) using the obtained continuum normalized 
spectra. For estimating these parameters, ZASPE compares the observed spectrum with a grid of 
synthetic stellar atmospheric models using least squares minimization. 
The best fit atmospheric parameters as well as the projected rotational velocities ($v\sin{i}$) and radial 
velocities ($v_\mathrm{rad}$) of the EDD host stars yielded by this approach are listed in Table~\ref{tab:props}.
For J092521, the weighted averages of the parameters obtained from its two measurements are presented in the table.
J104416 has a wide companion star of similar brightness (Gaia~DR3~5366814012930376960, Sect.~\ref{sec:multiplicity}), 
that we also observed with FEROS (Tab.~\ref{tab:feroslog}) and whose parameters were determined according to the method 
described above. The value for the metallicity of J104416 in Table~\ref{tab:props} is the weighted average of the 
results for the two stars. 
In addition to the fundamental stellar properties, Table~\ref{tab:props} also shows alternative identifiers, astrometric 
and photometric data, as well as galactic space position and velocity 
of the targets. 

\begin{table*} 
\setlength{\tabcolsep}{1.3mm}                                                   
\begin{center} 
\caption{Properties of newly discovered EDD systems. References: 1 - \citet{cutri2013}, 2 - Gaia~DR3, 3 - \citet{hog2000}, 
           4 - \citet{cutri2003}, 5 - \citet{henden2016}, 6 - \citet{cbj2021}, 7 - This work.  \label{tab:props}}
\begin{tabular}{lcccccc}                                                     
\hline\hline
Parameters & J060917 & J071206 & J092521 & J104416 & J204315 & Ref. \\
\hline
\multicolumn{7}{c}{Identifiers} \\
\hline
AllWISE  & J060917.00$-$150808.5 &  J071206.54$-$475242.3 &  J092521.90$-$673224.8 &  J104416.70$-$451613.9 &  J204315.23$+$104335.3 & 1 \\ 
Gaia DR3 & 2993153951148830336 &  5508896788120942848 &  5247128354025136128 &  5366813974272886528 &  1751335969361640192 & 2 \\
TYCHO    &    -                &  -                   &      9196-2916-1     &           -          &        -             &  3 \\
2MASS    &   06091701$-$1508085  &  07120655$-$4752423    &   09252194$-$6732251   &  10441680$-$4516142    &  20431522$+$1043355    & 4 \\
\hline
\multicolumn{7}{c}{Astrometric data} \\ 
\hline  
R.A. (2000) &  06:09:17.011 &  07:12:06.544 &  09:25:21.973 &  10:44:16.798 &  20:43:15.223 & 2 \\
Decl (2000) &  $-$15:08:08.59 &  $-$47:52:42.36 &  $-$67:32:25.06 &  $-$45:16:14.36 &  $+$10:43:35.60 & 2 \\
{\bf $\mu_{\alpha} \cos{\delta}$} (mas) & $-$6.271$\pm$0.011 & $-$8.606$\pm$0.154 & $-$17.794$\pm$0.021 & $-$89.291$\pm$0.009 & $+$12.996$\pm$0.015 & 2 \\
$\mu_{\delta}$ (mas)   & $+$0.882$\pm$0.012 & $-$2.096$\pm$0.159 & $+$25.950$\pm$0.022 & $+$29.754$\pm$0.010 & $-$15.934$\pm$0.013 & 2 \\
$\pi$ (mas)            & 2.832$\pm$0.013 &  5.751$\pm$0.125 &  10.106$\pm$0.020 &  5.007$\pm$0.011 &  8.479$\pm$0.014 & 2 \\
\texttt{ruwe}          & 0.96 &    12.27 &     1.24 &     0.91 &     0.99 & 2 \\
\hline
\multicolumn{7}{c}{Photometric data} \\ 
\hline 
$B$         &  14.34$\pm$0.05 &  13.92$\pm$0.05 &  13.04$\pm$0.04 &  12.94$\pm$0.06 &  13.27$\pm$0.04 &  5 \\ 
$V$         &   13.54$\pm$0.05 &  12.85$\pm$0.01 &  11.92$\pm$0.02 &  12.20$\pm$0.01 &  12.26$\pm$0.05 & 5 \\
$G_{BP}$    &   13.7431$\pm$0.0039 &  13.0773$\pm$0.0035 &  12.1514$\pm$0.0032 &  12.6394$\pm$0.0028 &  12.4620$\pm$0.0032 & 2 \\ 
$G_{RP}$         &   12.7325$\pm$0.0042 &  11.7441$\pm$0.0040 &  10.7075$\pm$0.0047 &  11.6415$\pm$0.0038 &  11.1882$\pm$0.0040 & 2 \\ 
$G$         &   13.3206$\pm$0.0029 &  12.5031$\pm$0.0030 &  11.5499$\pm$0.0028 &  12.2222$\pm$0.0028 &  11.8971$\pm$0.0028 & 2 \\ 
$J$         &   12.058$\pm$0.023 &  10.821$\pm$0.024 &  9.733$\pm$0.026 &  10.946$\pm$0.036 &  10.366$\pm$0.023 & 4 \\ 
$H$         &   11.643$\pm$0.022 &  10.296$\pm$0.027 &  9.174$\pm$0.026 &  10.540$\pm$0.034 &  9.824$\pm$0.028 & 4 \\ 
$K_{\rm s}$ &   11.481$\pm$0.023 &  10.176$\pm$0.025 &  9.004$\pm$0.024 &  10.460$\pm$0.030 &  9.679$\pm$0.023 & 4 \\
\hline
\multicolumn{7}{c}{Kinematics and positions} \\
\hline
$v_{\rm rad}$ (km~s$^{-1}$) & $+$23.44$\pm$2.32 & $+$27.64$\pm$1.10 &  - & $+$17.56$\pm$0.55 & $-$20.30$\pm$0.52 & 2 \\
              & $+$20.16$\pm$0.03 & $+$28.57$\pm$0.13 & $+$18.89$\pm$0.14 & $+$17.18$\pm$0.04 & $-$19.47$\pm$0.22& 7 \\
$U$ (km~s$^{-1}$)  & $-$16.34$\pm$0.03 & $-$6.02$\pm$0.13 & $-$9.06$\pm$0.05 & $-$82.26$\pm$0.17 & $-$9.45$\pm$0.11 & 7 \\ 
$V$ (km~s$^{-1}$)  & $-$7.17$\pm$0.04 & $-$24.61$\pm$0.14 & $-$22.15$\pm$0.13 & $-$35.92$\pm$0.06 & $-$20.25$\pm$0.16 & 7 \\ 
$W$ (km~s$^{-1}$)  & $-$14.05$\pm$0.05 & $-$15.06$\pm$0.19 & $-$1.43$\pm$0.03 & $-$11.84$\pm$0.03 & $-$3.99$\pm$0.07 &  7 \\ 	   
$X$ (pc)   & $-$248.69$\pm$1.26 & $-$32.55$\pm$0.67 & $+$25.22$\pm$0.05 & $+$36.01$\pm$0.07 & $+$61.59$\pm$0.10 & 7 \\ 
$Y$ (pc)   & $-$224.11$\pm$1.13 & $-$163.87$\pm$3.40 & $-$93.32$\pm$0.17 & $-$191.37$\pm$0.38 & $+$92.48$\pm$0.15 & 7 \\ 
$Z$ (pc)   & $-$96.20$\pm$0.49 & $-$49.57$\pm$1.03 & $-$20.75$\pm$0.04 & $+$41.50$\pm$0.08 & $-$38.19$\pm$0.06 &  7 \\

Distance (pc)      & 348.3$^{+1.8}_{-1.7}$ &  174.3$^{+3.2}_{-4.1}$ &  98.9$\pm$0.2 &  199.1$^{+0.3}_{-0.5}$ &  117.5$\pm$0.2 & 6 \\
\hline
\multicolumn{7}{c}{Stellar properties} \\
\hline
Spectral type &   G9V &  K3.5V &  K4.5V &  K0.5V &  K4V & 7 \\ 
$T_\mathrm{eff}$ (K)  &   5409$\pm$52 &  4682$\pm$69 &  4513$\pm$49 &  5241$\pm$68 &  4642$\pm$59 & 7 \\ 
$[\mathrm{Fe/H}]$ (dex) & $-$0.02$\pm$0.03 & $-$0.20$\pm$0.06 & $-$0.23$\pm$0.04 & $-$0.35$\pm$0.05 & $-$0.15$\pm$0.05 &  7 \\
$\log{g}$ (dex)         & 4.56$\pm$0.08 &  4.48$\pm$0.13 &  4.43$\pm$0.11 &  4.48$\pm$0.08 &  4.48$\pm$0.11 & 7 \\
$E(B-V)$                & 0.058$\pm$0.017 &     0.025$\pm$0.014    &     0.0$\pm$0.015     &   0.010$\pm$0.014  &  0.0$\pm$0.016 & 7 \\ 
$L_*$ (L$_\odot$)       &  0.493$\pm$0.019 &  0.284$\pm$0.017 &  0.224$\pm$0.010 &  0.406$\pm$0.021 &  0.207$\pm$0.011 &  7 \\
$R_*$ (R$_\odot$)       &  0.80$\pm$0.02 &  0.84$\pm$0.04 &  0.81$\pm$0.02 &  0.77$\pm$0.03 &  0.70$\pm$0.03 &  7 \\
$M_*$ (M$_\odot$)       & 0.90$\pm$0.01  & 0.69$\pm$0.02  & 0.65$^{+0.03}_{-0.02}$  & 0.83$\pm$0.02 & 0.72$\pm$0.01 &  7 \\
$v \sin{i_*}$ (km~s$^{-1}$) &  2.13$\pm$0.70 &  3.68$\pm$0.89 &  0.90$\pm$0.76 &  0.70$\pm$0.73 &  3.18$\pm$0.93 & 7 \\ 
$P_{\rm rot}$ (day)       & 6.55$\pm$0.08  & 8.55$\pm$0.17  & 4.95$\pm$0.05  & - & 9.0$\pm$0.2 &  7 \\ 
Li\,I EW (\,m\AA)        & 143$\pm$5   & $<$21  & 108$\pm$2  & $<$11 & 35$\pm$4  &  7 \\          
Age (Myr)        & 208$^{+104}_{-65}$  & 243$^{+91}_{-77}$  & 85$^{+49}_{-28}$  & 5500$^{+2400}_{-2200}$ & 252$^{+80}_{-65}$ &  7 \\
Multiplicity     &  Yes  & Likely  & ?  & Yes  & Yes  &  7 \\
\hline
\multicolumn{7}{c}{Disc properties} \\
\hline
$T_\mathrm{d}$ (K)  & 752$\pm$48 / 237$\pm$22$^a$ & 434$\pm$25  & 857$\pm$90 / 99$\pm$6$^a$  & 405$\pm$21 & 404$\pm$15 &  7 \\
$f_\mathrm{d}^b$   & 0.075/0.027$^a$  & 0.026  & 0.021/0.033$^a$  & 0.025  & 0.048 &  7 \\ 
$R_\mathrm{BB}$ (au)  & 0.1/0.97  & 0.22  & 0.05/3.7  & 0.30 & 0.22  &  7 \\
mid-IR var.   & Y  & N  & Y  & N & Y &  7 \\  
\hline
\multicolumn{7}{l}{$^a$ Spectral energy distributions of excess emission of J061709 and J092521 are fitted with two-temperature 
blackbody models (Sect.~\ref{sec:excessmodel}).}\\
\multicolumn{7}{l}{$^b$ $f_d$ is the fractional luminosity ($f_d = L_\mathrm{disc} / L_*$).} \\
\\
\end{tabular}
\end{center}
\end{table*}

\subsection{Equivalent width of the Li\,I line}
Using the continuum normalized FEROS spectra we determined the equivalent width of the Li\,I line.
To this end, the two Li\,I components at 6707.761 and 6707.912\,{\AA} and the very nearby Fe\,I line 
at 6707.431\,{\AA} were fitted as a combination of three Gaussians. The obtained EWs are listed in Table~\ref{tab:props}.

\subsection{Multiplicity} \label{sec:multiplicity}
Data available in the Gaia DR3 catalogue, supplemented by our 
spectroscopic results, allow us to search for possible companion stars of our targets 
located at different separations, from close spectroscopic binaries to the widest 
pairs. By examining cross correlation function (CCF) profiles calculated from our 
high resolution spectra, we found no indication that any of the host stars of 
are double-lined spectroscopic binaries. RV data for all targets, except J092521, 
are available in the Gaia DR3 catalogue and are in good agreement with the radial 
velocities obtained from FEROS observations (the RV differences are in all cases less 
than 1.5$\times$ the uncertainty of the difference, see Table~\ref{tab:props}). For J092521, 
we have FEROS data for two epochs (Table~\ref{tab:feroslog}) 
and the RV velocities obtained are in perfect agreement.
The RV values reported in DR3 catalogue are a combination of all individual measurements 
performed in the first 34 months of the observing mission \citep{katz2023}.
Although the measured radial velocity time series are currently 
published only for a limited number of objects, the catalogue already includes 
two indices \citep[\texttt{rv\_chisq\_pvalue} and \texttt{rv\_renormalised\_gof},][]{katz2023} 
that can be used to identify RV variability for stars with 
3900\,K $<$ \texttt{rv\_template\_teff} $<$ 8000K, that are sufficiently bright 
(\texttt{grvs\_mag} $<$ 12\,mag) and have been observed at least ten times. 
Since these indices became available in the DR3 release, we investigated possible RV variability 
for all systems targeted in our project, with the exception of J092521, for which there are no RV data 
in the catalogue.
We found that TYC\,5940 fulfills the criteria (\texttt{rv\_chisq\_pvalue}$<$0.01 and 
\texttt{rv\_renormalised\_gof}$>$4) recommended by \citet{katz2023} and thus exhibits significant 
RV variability raising the possibility that it has a close companion. 
However, we note that our ground-based radial velocity measurement for the star \citep[$+$28.50$\pm$0.20~km~s$^{-1}$,][]{moor2021} 
is in good agreement with the mean RV data listed in the Gaia\,DR3 catalogue ($+$27.82$\pm$1.05~km~s$^{-1}$), 
which does not indicate any significant change.

The re-normalized unit weight error (\texttt{ruwe}) 
indicates the quality of the applied astrometric solution that is 
based on a single-star model \citep{lindegren2021}. A \texttt{ruwe} value greater than 1.4 implies a potentially 
unreliable astrometric solution \citep{fabricius2021}, which in most cases, although not exclusively 
\citep[e.g.,][]{fitton2022}, may be due to the presence of an unresolved close 
companion star located within $\lesssim$1\arcsec. In fact, according to several 
works \citep[e.g.,][]{belokurov2020,bryson2021,wolniewicz2021}, unresolved binarity 
is suspected already at \texttt{ruwe} above 1.2. In the light of these results, 
the high \texttt{ruwe} of 12.3 for J071206 is a strong indication that the star 
may be an unresolved pair, while the \texttt{ruwe}=1.24 listed for J092521 suggests 
only tentatively the presence a companion. However, for the latter
the large value of the \texttt{ipd\_frac\_multi\_peak} metric (21, Gaia~DR3) -- that measures 
what fraction of observing windows where multiple peaks are detected -- further strengthens 
that the source may not be single \citep{fabricius2021}.  

To reveal possible wide companions of the targets we looked for co-moving and co-distant 
stars in their vicinity using the method proposed by \citet{deacon2020}. Only objects with 
\texttt{ruwe}$<$1.4 and $\pi/\sigma_\pi > 4$ (where $\pi$ and 
$\sigma_\pi$ are the measured parallax and its uncertainty) were considered in the search, 
and we set a maximum projected separation of 30000\,au for the possible companions to avoid 
binary candidate likely to be chance alignments only \citep{elbadry2021}. 
We have found potential wide companions to three stars, J060917, J104416, and J204315.  
The most important Gaia astrometric parameters along with the projected 
separations of these pairs are summarised 
in Table~\ref{tab:companions}. We note that for the target Gaia~DR3~5366814012930376960, the pair of J104416, 
radial velocity is also available in the catalogue, and the quoted RV value of $17.12\pm0.37$\,km~s$^{-1}$ 
is in very good agreement 
with that of J104416 (Table~\ref{tab:props}), further confirming that the two stars constitute a binary.
While for J060917 and J204315, the companions are fainter than the discs' host stars, 
Gaia~DR3~5366814012930376960 is brighter than J104416, implying that in this system the revealed companion is 
the primary component.   
The very faint J060917\,B is not, but the other two pairs are listed in the wide binary catalogue compiled by \citet{elbadry2021}.

For Gaia~DR3~5366814012930376960 we used the ZASPE code to derive its fundamental stellar properties
from its high-resolution FEROS spectrum (Sect.~\ref{sec:stellarprops}) and obtained the following results: 
$T_\mathrm{eff}$ = 5497$\pm$78\,K, $\log{g}$ = 4.43$\pm$0.13, [Fe/H] = -0.26$\pm$0.04 (weighted average of the [Fe/H] 
estimates for Gaia~DR3~5366814012930376960 and J104416), $v \sin{i}$ = 1.77$\pm$0.85\,km~s$^{-1}$, and $v_\mathrm{rad}$ = 
17.06$\pm$0.01\,km~s$^{-1}$. 
We derived a luminosity of 0.575$\pm$0.033\,L$_\odot$ for the star and $<$10\,{m\AA} for its lithium EW.

To estimate the effective temperature and spectral type of J204315\,B, we used its 
$r-z$, $r-J$, and $G_{BP} - G_{RP}$ colour indices and the method proposed by \citet{amann2015}.  
The $r$ and $z$ band photometry were taken from the Pan-STARRS Survey \citep[DR1;][]{chambers2016} 
and were transformed into the SDSS system utilizing the equations presented in \citet[DR1;][]{tonry2012}. 
The $J$ band data were extracted from the 2MASS survey \citep{skrutskie2006}, while $G_{BP} - G_{RP}$ 
colours were from the Gaia\,DR3 catalogue. 
Since J204315 has negligible reddening (Sect.~\ref{sec:photmodels}), in this analysis 
we adopted E(B-V) = 0 for 
the companion too.
The obtained $T_\mathrm{eff}$ was then used to estimate the spectral type of J204315\,B based on the online database\footnote{\url{https://www.pas.rochester.edu/~emamajek/EEM\_dwarf\_UBVIJHK\_colors\_Teff.txt}} of 
stellar parameters of dwarf stars \citep{pecaut2013}. 
The spectral type of J060917\,B, where reliable photometry was available only in the $G$ band, was estimated 
based on its $M_G$ absolute magnitude by interpolating in the same database. 
In estimating the parameters of J204315\,B and J060917\,B, the distance of their 
primary component has been adopted (Table~\ref{tab:props}).
The derived stellar parameters of the companions are quoted in Table~\ref{tab:companions}.

\begin{table*}                                                                  
\setlength{\tabcolsep}{1.1mm}                                                   
\begin{center}                                                                                     
\caption{Astrometric parameters and stellar properties of the probable common motion and distance pairs. \label{tab:companions} }
\begin{tabular}{lccccccccc}							  
\hline\hline
Gaia DR3 id &   Pair	  & Separation    & {\bf $\mu_{\alpha} \cos{\delta}$} & $\mu_\delta$ & $\pi$ & $G$   &  SpT & $T_\mathrm{eff}$ & $M_G$ \\
	   &		  &  (\arcsec/au) &   (mas)	   &   (mas)	  & (mas) & (mag) &      &   (K)  	       & (mag) \\	
\hline
2993153951145272192 & J060917 & 1\farcs96/682 &   $-$4.421$\pm$0.589 &  $+$0.554$\pm$0.721  & 3.527$\pm$0.775 & 20.103$\pm$0.009  & M4.5 & - & 12.286$\pm$0.039 \\  
5366814012930376960 & J104416 & 51\farcs56/10261 &$-$89.165$\pm$0.011 & $+$29.954$\pm$0.012 & 4.995$\pm$0.013 & 11.843$\pm$0.003  & G8V  & 5497$\pm$78 & 5.295$\pm$0.027 \\
1751335969361640576 & J204315 & 18\farcs96/2228 & $+$12.556$\pm$0.039 & $-$16.459$\pm$0.039 & 8.508$\pm$0.037 & 15.733$\pm$0.003  & M3.5 & 3244$\pm$48 & 10.383$\pm$0.016 \\
\hline
\end{tabular}
\end{center}
\end{table*}

\subsection{Stellar photosphere models} \label{sec:photmodels}
To estimate the stellar luminosity and the reddening as well as to predict the photospheric contribution to the 
total flux at IR wavelengths, the optical and near-IR spectral energy distributions (SEDs) of the stars were fitted by ATLAS9 atmosphere models \citep{castelli2003}. The broadband photometric data, utilized in the modelling, were taken  
from the APASS \citep[AAVSO Photometric All-Sky Survey;][]{henden2016}, TYCHO2 \citep{hog2000}, Gaia\,DR3, and 
2MASS \citep[Two Micron All Sky Survey;][]{cutri2003} catalogues. From the latter data set, only the $J$- and $H$-band photometry was used, as because of the presence of warm dust, the possibility of some IR excess 
even in the $K$-band cannot be excluded. In the case of J092521, the Gaia $B_\mathrm{P}$ and $R_\mathrm{P}$ 
data were also discarded, because the corrected $B_\mathrm{P}$ and $R_\mathrm{P}$ flux excess factor \citep{riello2021} 
suggests a significant inconsistency between photometry in these bands with respect to 
$G$-band data. To account for possible interstellar reddening, 
we adopted the extinction law from \citet{fitzpatrick1999}. In the course of fitting, we used a grid-based approach. 
In the first step, the $\log{g}$ and [Fe/H] parameters of the atmospheric model were fixed to the best-fit values derived 
in our spectroscopic analysis but $T_\mathrm{eff}$ was free to vary. For J060917, J104416, and J204315, the obtained best fit 
$T_\mathrm{eff}$ values were in agreement with the estimates derived from the spectral analysis within 30\,K. 
In these cases, 
we finally fixed the effective temperatures to the latter values and repeated the photospheric fitting. For J071206 and J092521, the fitting resulted in models with temperatures of 80 and 90\,K lower, respectively, than the ones obtained from the analysis of their spectra. One possible explanation for this is that these two objects may have fainter, cooler companions -- as suggested by their Gaia data (Sect.~\ref{sec:multiplicity}) -- that 
also contribute to the total measured flux in all available photometric bands. For these sources, therefore, we kept the 
resulting colder photospheric models. The luminosities ($L_*$) and reddenings (E(B-V)) obtained by the fitting 
are given in Table~\ref{tab:props}. We note that the resulting E(B-V) values are in good agreement with 
the ones that can be extracted for our targets from the Stilism 3D reddening map\footnote{\url{https://stilism.obspm.fr}} \citep{capitanio2017,lallement2018}.

\subsection{Optical variability} \label{sec:opticalvar}

Our targets are G9--K4.5 type stars, they are likely magnetically active, thus, they may be spotted, which leads to rotational 
modulation in their light curves. To search for possible modulation and to determine the stellar rotational periods, we studied 
the optical variability of our targets 
using photometric observations obtained by the Transiting Exoplanet Survey Satellite \citep[TESS;][]{ricker2015} and 
in the ground based All-Sky Automated Survey for Supernovae \citep[ASAS-SN;][]{shappee2014,kochanek2017} project.
The data taken by ASAS-SN are also useful to investigate possible longer-term trend-like variations in the brightness of stars, which may be 
important in determining whether the variability seen in the mid-IR is related to discs or not (Sect.~\ref{sec:irvar}).  

\subsubsection{TESS observations}
All five EDD host stars in Table~\ref{tab:props}, as well as Gaia~DR3~5366814012930376960, the companion of J104416,
were observed by the TESS
in the full-frame image (FFI) mode, enabling us to investigate their variations on daily/weekly timescales, including rotational modulations, 
with a high precision. Table~\ref{tab:tesslog} shows the TESS Input Catalog (TIC) identifiers of our targets along with the information 
in which TESS Sectors they were measured. While during the TESS primary mission (S1-S26) the full frame images were acquired at a 
30-min cadence, later, in the extended mission a shorter 10-min cadence was used. In our analysis, we used un-detrended 
light curves extracted by the MIT Quick Look Pipeline \citep{huang2020a,huang2020b,kunimoto2021,kunimoto2022} from the FFI images using simple aperture photometry.
These data were downloaded from the Mikulski Archive for Space Telescopes (MAST\footnote{\url{https://mast.stsci.edu}}).
J204315 was only observed in one sector, and 46\% of its light curve are flagged as having bad quality by the QLP pipeline 
(because of straylight from Earth or Moon in camera FOV and/or low precision). Therefore, this measurement was not included 
in the analysis. To search for periodic changes in the other light curves we constructed their periodograms
using the Generalized Lomb-Scargle algorithm \citep[GLS;][]{zechmeister2009} utilizing its implementation in 
PyAstronomy\footnote{\url{https://github.com/sczesla/PyAstronomy}} \citep{pya}. The periodograms were calculated 
separately for each $\sim$27\,d long sectors. No significant periodic variation was found for J104416 and Gaia~DR3~5366814012930376960.
J071206 and J092521 show periodic modulation in all six sectors in which they were observed. We selected the most 
significant peaks 
in each sector, and the final period and its uncertainty was computed as the average and standard deviation of the 
individual results. This approach yielded periods of 8.55$\pm$0.17 and 4.95$\pm$0.05\,d for J071206 and J092521, 
respectively. For J060917, we also found a significant peak at 6.55$\pm$0.08\,d in the periodogram. 
The amplitudes of the detected modulations range between 0\fm001 and 0\fm008. 

\subsubsection{ASAS-SN observations}
Our targets were also covered by the ASAS-SN providing $V$-band
photometry between 2012 and 2018 and $g$-band data since 2018. 
To compile the light curves, we used the \textsc{Sky Patrol} tool\footnote{\url{https://asas-sn.osu.edu/}}. 
After removing longer term trends from these data using a LOWESS algorithm, 
we calculated GLS periodograms for the detrended light curves. For J071206, a significant peak was found in the 
$V$ and $g$ band measurements with periods of 8.67 and 8.61\,d respectively, which is in good agreement with the 8.55\,d period found in the TESS dataset. The case of J204315 is similar, we found periodic signals in both light curves 
 with a period of 9.06$\pm$0.12\,d in the $V$-band and a period of 8.88$\pm$0.16\,d in the $g$-band. In our analysis we will use the weighted average of the two results is 9.0$\pm$0.1\,d. 
 
The derived rotational periods are listed in Table~\ref{tab:props}.
 
\begin{table}                                                                  
\begin{center} 
\caption{TESS observations \label{tab:tesslog}}
\begin{tabular}{lcl}                                                     
\hline\hline
Name & TIC identifier & Sectors \\
\hline
J060917 & TIC\,408967891 & S33  \\
J071206 & TIC\,134435629 & S6--8, S33--35 \\
J092521 & TIC\,304170889 & S9--11, S36-38 \\
J104416 & TIC\,146890996 & S9--10, S36  \\
Gaia~DR3~5366814012930376960 & TIC\,146890957 & S9--10,S36 \\
J204315 & TIC\,282627688 & S55  \\
\hline
\\
\end{tabular}
\end{center}
\end{table}

\begin{figure*}
    \centering
    \includegraphics[width=0.49\textwidth]{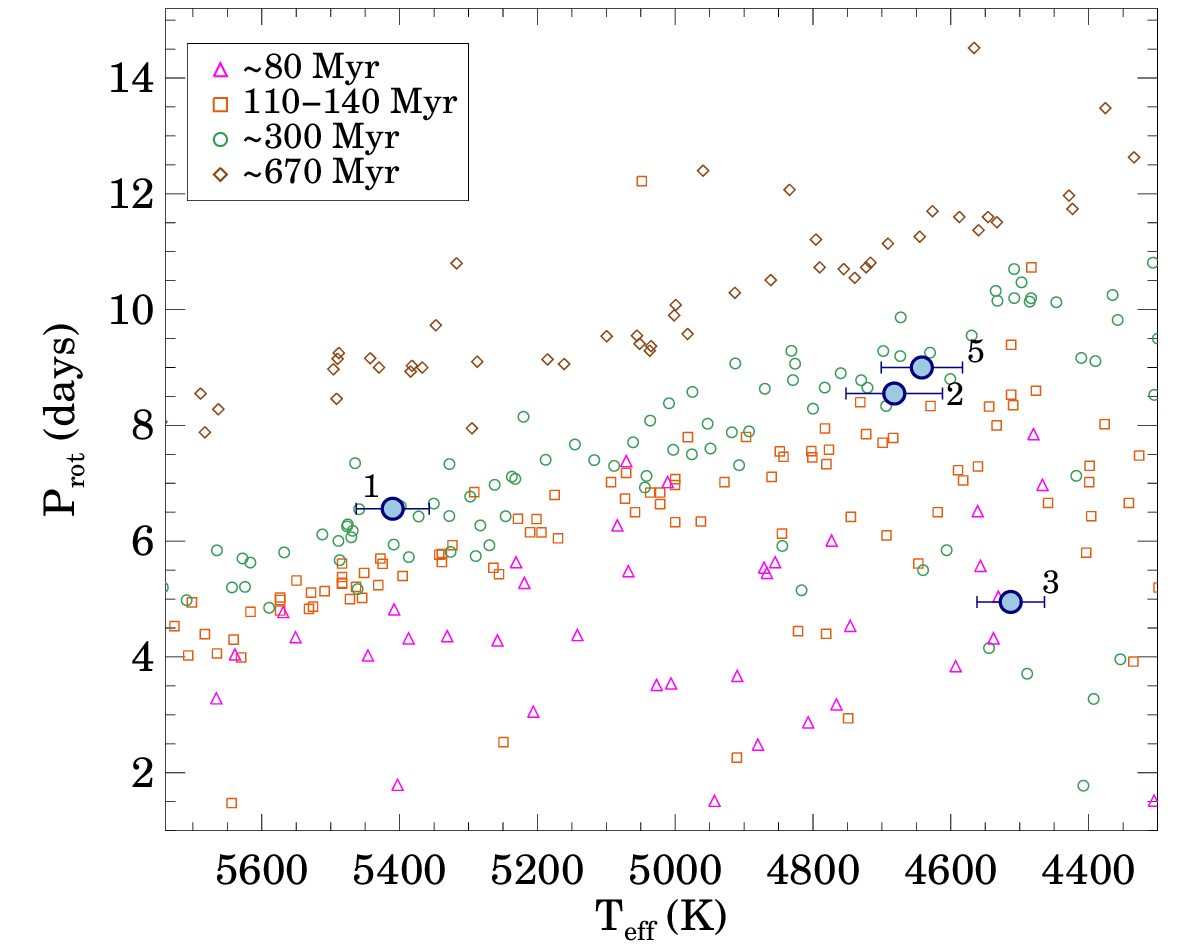}
    \includegraphics[width=0.49\textwidth]{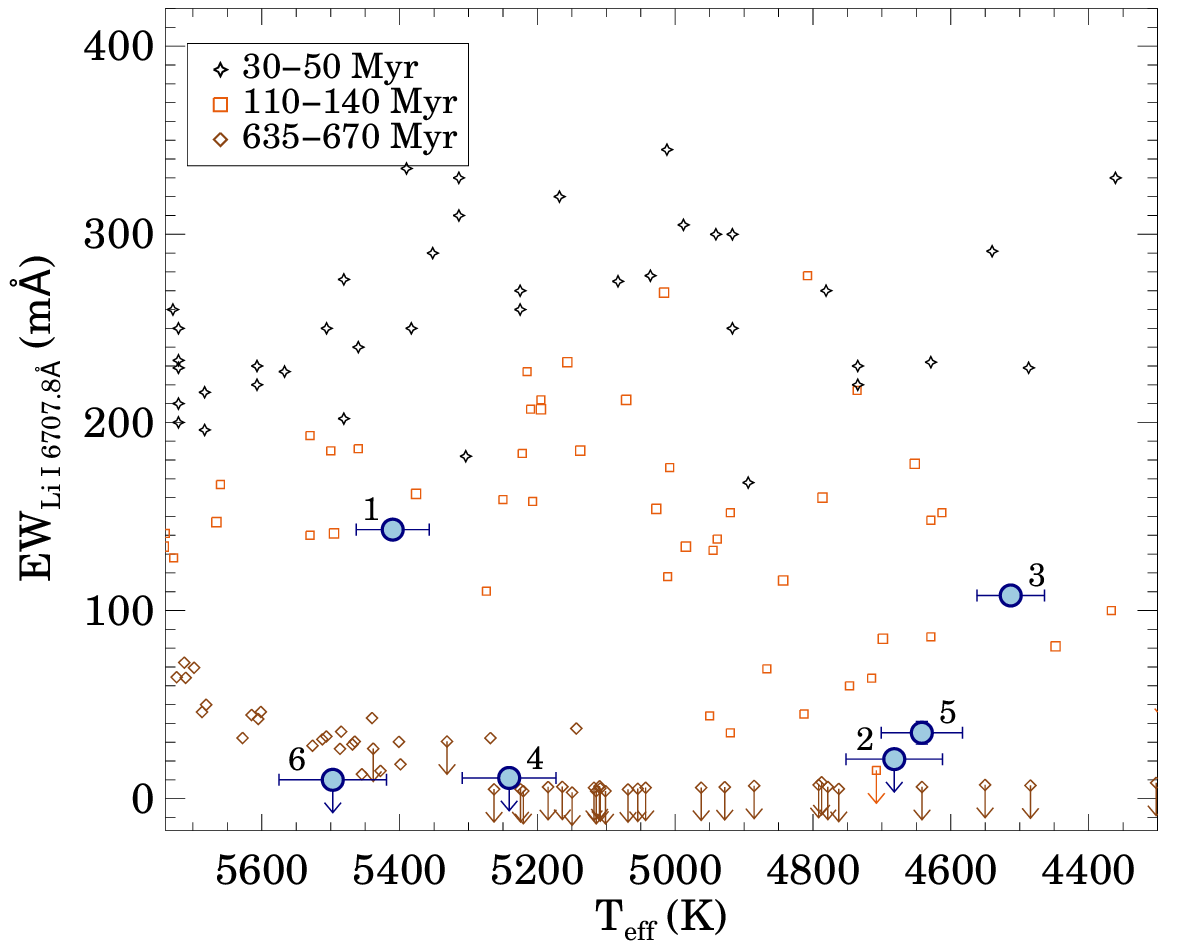}
    \caption{Left: Rotation periods as a function of effective temperatures for four of our targets (filled blue circles with error bars)
and for members of seven stellar groups: $\alpha$~Per ($\sim$80\,Myr), Blanco-1, Pleiades, Psc-Eri (110-140\,Myr), NGC\,3532, Group-X ($\sim$300\,Myr), and Praesepe ($\sim$670\,Myr). Only members that are likely to be single and have a reliable rotation period were plotted. All data were taken from \citet{bouma2023}. Right: Equivalent widths of the Li\,I 6708\,{\AA} as a function of 
effective temperature measured for our targets. For comparison Li\,I data of members of 8 stellar groups are also plotted: 
 Tucana-Horologium, Carina, Columba, and Argus moving groups \citep[30--50\,Myr,][]{dasilva2009}; Blanco-1 and Pleiades open 
 clusters \citep[110-140\,Myr,][]{sestito2005,bouvier2018}, Praesepe and Hyades open clusters \citep[635-670\,Myr,][]{cummings2017}.
Our targets are marked by the following numbers in the plots: 
1: J060917, 2: J071206, 3: J092521, 4: J104416, 5: J204315, 6: Gaia~DR3~5366814012930376960.  
    }
    \label{fig:age}
\end{figure*}

\subsection{Stellar ages}
Our age estimates for stars are mainly based on the rotation-age \citep[gyrochronology,][]{barnes2003} and 
lithium depletion-age relationships, although in some cases kinematic information and position on the HRD were 
also considered. Figure~\ref{fig:age} shows the rotation periods (left panel) -- where available
(Sect.~\ref{sec:opticalvar}) -- and the lithium equivalent width (right panel) of the stars as a function of effective temperature.   
Similar data of single stars belonging to nearby moving groups and open clusters of different ages
are also plotted. A comparison with these well-dated objects gives an immediate rough estimate on the age of our 
stars. To quantify the best age estimate and its uncertainty, we applied the \textsc{gyro-interp} \citep[][]{bouma2023} and \textsc{eagles} \citep[Estimating AGes from Lithium Equivalent widthS,][]{jeffries2023} tools. Both packages use empirical models calibrated with measurements of stars in clusters of known age and provide posterior age probability distributions.
The \textsc{gyro-interp} uses the effective temperature and the rotation period, while the \textsc{eagles} the effective temperature and the lithium  equivalent width as input parameters.
The obtained posteriors were multiplied and the median age and the 68\% confidence interval were calculated
using the final distribution.

\begin{figure}
\includegraphics[width=0.49\textwidth]{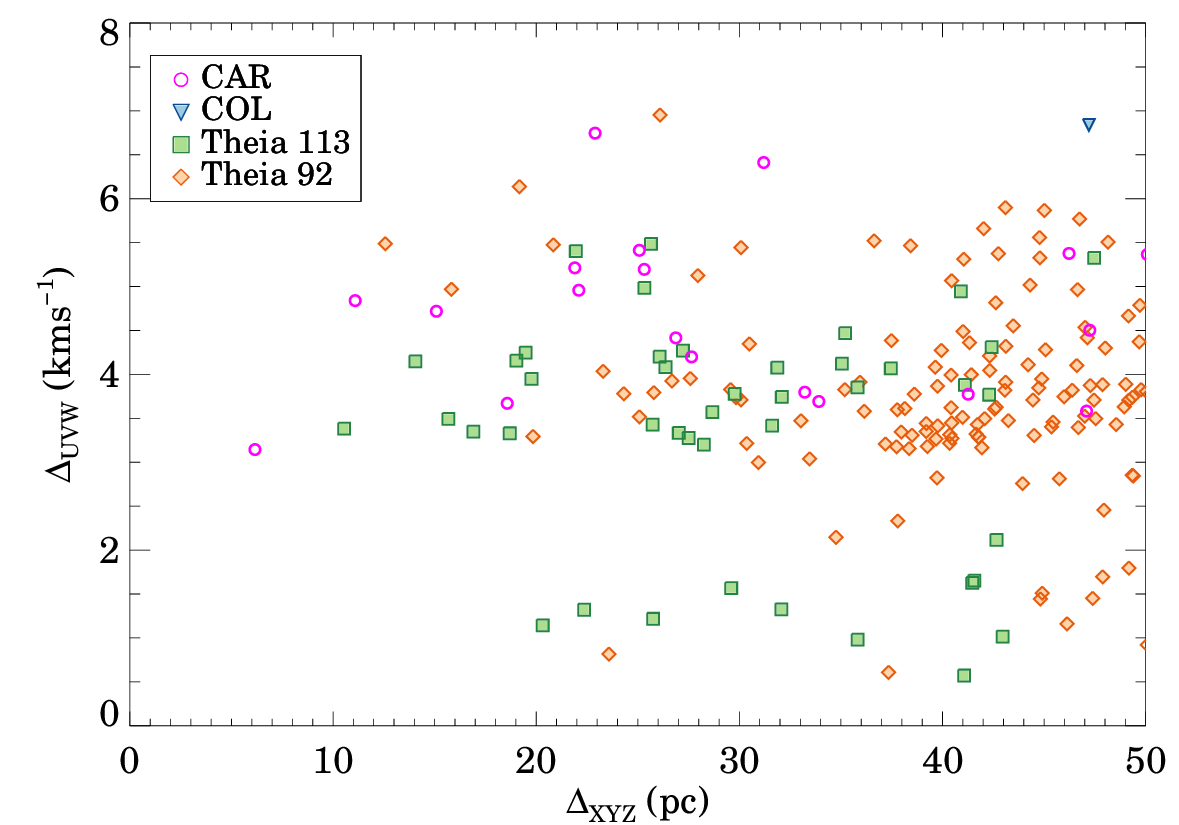}
\caption{Velocity ($\Delta_\mathrm{UVW}$) and positional differences ($\Delta_\mathrm{XYZ}$) of members 
of five young stellar groups with respect to J092521. Member lists of the groups are taken from the literature: 
Carina (CAR) - \citet{booth2021}, Columba (COL) - \citet{gagne2018a,gagne2018b,torres2008}, Theia\,92 and Theia\,113 - \citet{kounkel2019}.
Note that in some cases the same star 
has been assigned to more than one group.}
\label{fig:j092521}
\end{figure}

For J060917 and J204315, the combined posteriors yielded age estimates of 208$^{+104}_{-65}$\,Myr and 252$^{+80}_{-65}$\,Myr,
respectively. The latter star has a wide companion at an angular separation of $\sim$19{\arcsec} (Sect.~\ref{sec:multiplicity}). 
We compared the position of this M3.5-type star with those of M-type members of various well-dated associations 
and open clusters on a $M_\mathrm{G}$ versus $G-R_\mathrm{p}$ colour-magnitude diagram provided by 
\citet{popinchalk2021} (their figure~8). The figure clearly distinguishes M-type stars younger than 50\,Myr from those 
between 100 and 750\,Myr, with J204315~B falling into the latter sample, consistently with our age estimate for 
the primary component.

In the case of J071206, the obtained upper limit on the lithium equivalent width indicates a 95 per cent lower  
limit of 250\,Myr for its age. For this star, the high \texttt{ruwe} value and the high luminosity relative to its temperature indicate the presence of a companion that is likely located within  $\sim$1{\arcsec}. 
Assuming that the observed periodic modulation is related to the rotation of J071206 and not to the putative fainter companion, 
the \textsc{gyro-interp} tool suggests a slightly younger age of 230$^{+100}_{-70}$\,Myr. 
By combining the two posteriors we obtained a final estimate of 243$^{+91}_{-77}$\,Myr. Interestingly, 
the galactic space location and motion of J071206 resembles that of the recently revelead Crius\,228 group 
\citep{moranta2022}. Moreover, the estimated age of the group (231$\pm$35\,Myr) is consistent with that of the 
star (Appendix:\ref{sec:appendix}). Together, these seem to be a strong indication for group membership, but it should be kept in mind that the high \texttt{ruwe} also implies more uncertain astrometric parameters. 
Future Gaia data releases
may be able to separate the two components or provide an astrometric solution considering the likely non-single nature 
of this system. Therefore, in the present work we will use the age derived from the stellar data alone.

\begin{figure*}
\centering
\includegraphics[width=0.96\textwidth]{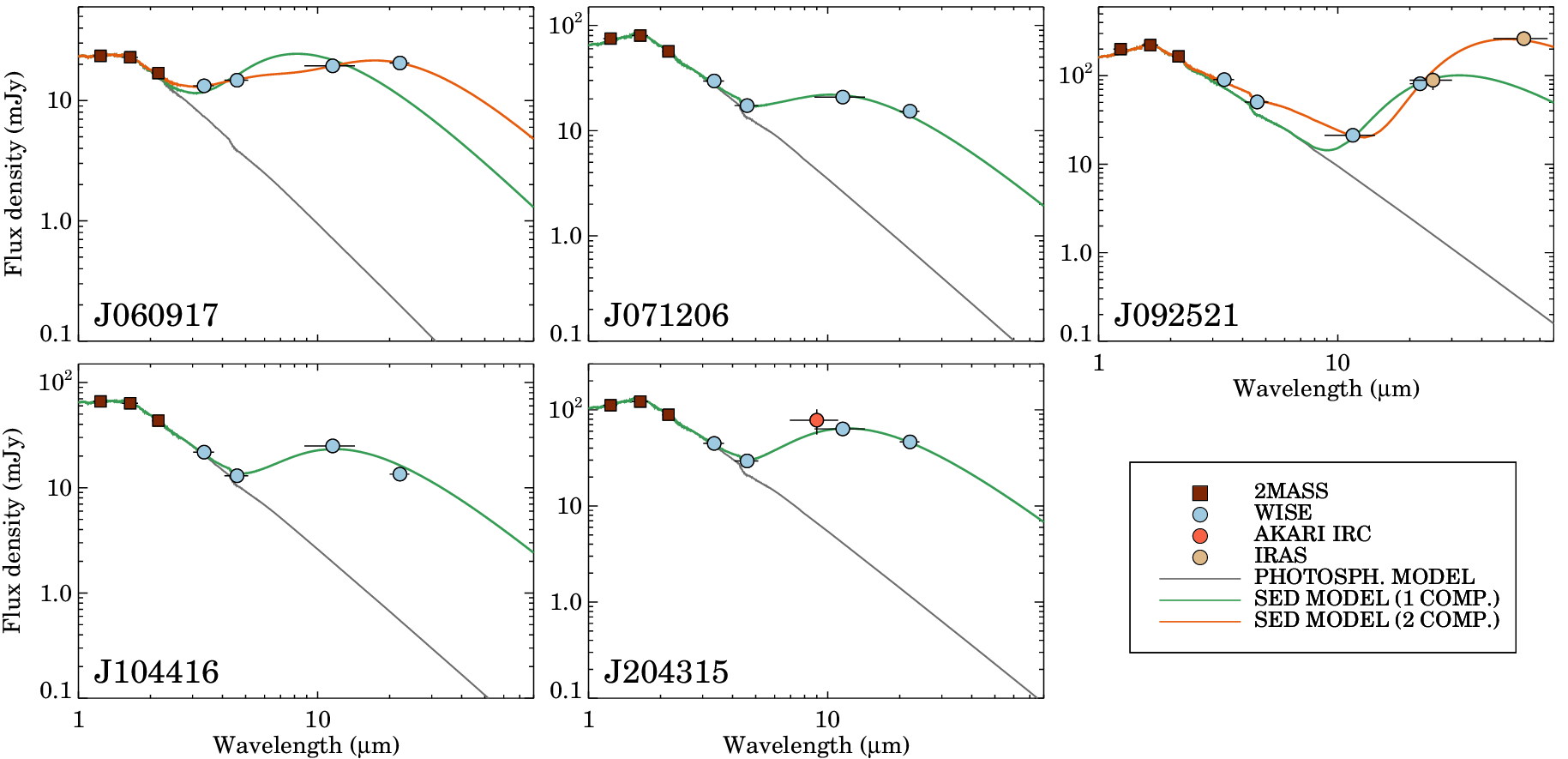}
\caption{Spectral energy distribution of the four newly discovered EDDs and J092521 with the photosphere models 
and the best fit SED models. The data points shown are dereddened (using the E(B-V) estimates from Table~\ref{tab:props}) 
and colour corrected. Horizontal bars show the width of the filters used to take the photometric points.}
\label{fig:diskmodels}
\end{figure*}

J092521 is the youngest system in the sample, with its lithium and rotation data together giving an age estimate of 
85$^{+49}_{-28}$\,Myr. It is also worth evaluating whether the star belongs to any known young kinematic assemblages.
Based on the Gaia DR3 astrometric parameters and our radial velocity measurement (Table~\ref{tab:props}), the \textsc{BANYAN $\Sigma$} online tool\footnote{\url{https://www.exoplanetes.umontreal.ca/banyan/banyansigma.php}} \citep{gagne2018a} -- 
which is designed to identify possible members in 27 nearby young stellar associations -- found a membership 
probability of 13\% to the $\sim$15\,Myr old Lower Centaurus Crux (LCC) subgroup of the Scorpio-Centaurus 
association and 87\% to field. On top of the low probability obtained in the \textsc{BANYAN $\Sigma$} model, there are further counter-arguments to LCC membership: our estimate indicates a substantially older age and in accordance with this, 
the star's luminosity is 30\% lower than even the faintest K-type LCC member with similar temperatures \citep[4400--4600\,K;][considering the new Gaia\,DR3 based distances of the members in the luminosity calculations]{pecaut2016}. 
\citet{higashio2022} also argues that J092521 is a young star and noted that while its velocity
is  more consistent with membership in the Tucana-Horologium association, its position is closer to the 
Columba and Carina moving groups. With regard to the latter two groups, it is worth noting that, as 
\citet{gagne2021} has recently proposed, they, together with Theia\,113 and Theia\,208 groups and part of 
Theia\,92 \citep[the latter three are identified by][]{kounkel2019}, may 
constitute the tidal tail of the Platais\,8 open cluster. To study the possible membership of J092521 further, we collected the members of 
the abovementioned groups from the literature \citep{booth2021,gagne2018a,gagne2018b,kounkel2019,torres2008,zerjal2023} and computed the UVW velocities and XYZ space positions of those stars that have reliable astrometric data (\texttt{ruwe}$<$1.4) and  radial velocities in the 
Gaia~DR3 catalog. 
Following \citet{gagne2021}, to avoid possible contaminations, in the cases of Theia groups only
member candidates separated by $<$5\,km~s$^{-1}$ from their group characteristic velocity in the UVW space
were kept.
Figure~\ref{fig:j092521} shows the positional $\left(\Delta_\mathrm{XYZ} = \sqrt{ (X-X_0)^2 + (Y-Y_0)^2 + (Z-Z_0)^2}\right)$, 
and velocity differences $\left(\Delta_\mathrm{UVW} = \sqrt{ (U-U_0)^2 + (V-V_0)^2 + (W-W_0)^2}\right)$ of these 
stars from J092521 whose data ($X_0,Y_0,Z_0,U_0,V_0,W_0$) are taken from Table~\ref{tab:props}. 
The figure shows that both in terms of galactic space positions and velocities, members of 
Carina, Theia\,92 and Theia\,113 are the closest to J092521. 
Age estimates for the Carina moving group vary widely in the literature, with some recent studies leaning towards a 
younger age of 13--27\,Myr \citep{booth2021,kerr2022,schneider2019,ujjwal2020}, 
while \citet{bell2015} proposed an age of 45$^{+11}_{-7}$\,Myr, and the most recent work based on the Lithium Depletion Boundary 
method yielded an age of 41$^{+3}_{-5}$\,Myr \citep{wood2023}. 
For Theia\,92 and Theia\,113, the estimates range between 45 and 63\,Myr and 45 and 71\,Myr 
\citep{gagne2021,kounkel2020}, respectively. Based on the age of the groups, it is therefore more likely that 
J092521 belongs to Theia\,92 or Theia\,113 groups, but membership in the Carina cannot be completely excluded either.  

J104416 forms a wide binary with Gaia~DR3~5366814012930376960. We were unable to determine the rotation period for any 
of the components. Based on the measured upper limits of the lithium equivalent widths, the \textsc{eagles} provides 
age estimates of $>$970 and $>$1310\,Myr for J104416 and Gaia~DR3~5366814012930376960, respectively.
To refine this result, we performed isochrone fitting using stellar evolutionary tracks from 
the Dartmouth Stellar Evolution Program \citep[DSEP,][]{dotter2008} and the \textsc{kiauhoku} package developed by \citet{claytor2020}. 
This was done for both stars, using their luminosity, effective temperature, and metallicity data as input parameters.
Taking into account both age posteriors obtained as a result of these fits and the lower limit from the lithium 
depletion, a final posterior is calculated from which an age estimate of 5.5$^{+2.4}_{-2.2}$\,Gyr was obtained for the system.
Because of the large total space velocity ($\sqrt{U^2 + V^2 + W^2} \sim 90$\,km~s$^{-1}$), it is a question 
whether J104416 might be a thick disc star. However, based on the kinematic criteria proposed by \citet{reddy2006}, it definitely belongs to the thin disc population. 
Using the method described in \citet{almeida-fernandes2018} we obtained a kinematic age estimate of 
7.5$^{+4.3}_{-3.5}$\,Gyr from the UVW data of the star. This is consistent with our result based on 
lithium content and isochrone fitting and adds another argument that this system is quite old.

\subsection{Infrared excess and fundamental disc properties}
\subsubsection{Construction of the SEDs}
Figure~\ref{fig:diskmodels} displays the SEDs of the five systems.
Most of the data at wavelengths longer than 3\,{\micron} come from measurements obtained by the {\sl Wide-Field Infrared 
Survey Explorer} \citep[{\sl WISE},][]{wright2010}. During its mission, between 2010 and 2022, the {\sl WISE} spacecraft 
measured our targets in 20--21 observing windows, separated by $\sim$180\,d, except for a gap between 2011 and 2013 
due to hibernation of the satellite. Due to coolant depletion after 7\,months, data in most windows are limited to the
$W1$ and $W2$ photometric bands at 3.4 and 4.6\,{\micron}, the longer wavelength $W3$ (at 12\,{\micron}) and $W4$ (at 22\,{\micron}) observations are only available in the first two windows for J092521 and in the first window for the other sources. Figure~\ref{fig:diskmodels} shows the {\sl WISE} data obtained in the first observing window, in 2010.
To derive these data, we downloaded the photometric results of the single-exposure observations 
from AllWISE Multiepoch Photometry tables of the NASA/IPAC Infrared Science Archive (IRSA\footnote{\url{https://irsa.ipac.caltech.edu/applications/wise/}}). After discarding bad quality data points, 
using the same strategy as in \citet{moor2021}, we computed the average of the remaining values.

We used the \textsc{SCANPI\footnote{\url{https://irsa.ipac.caltech.edu/applications/Scanpi/}}} tool to check if any of the sources were detected in the all-sky  survey performed by the {\sl The Infrared Astronomical Satellite} \citep[{\sl IRAS},][]{neugebauer1984}. 
We found that J092521 was detected both at 25 and 60\,{\micron}. This source is also listed in the 
IRAS Faint Source Reject Catalog \citep{moshir1992} under the identifier of IRAS\,Z09244-6719. The 25 and 60\,{\micron} 
flux densities reported there, 78.1$\pm$17.6\,mjy and 267.5$\pm$39.3\,mJy, respectively, are in good agreement with the SCANPI results, so we finally used the catalogue data. Considering the proper motion of J092521, taken from the Gaia DR3 catalogue, 
the star is located 8\farcs4 from the IR source, within the error ellipse. The {\sl WISE} ~$W4$-band image shows that there are no 
other bright sources in the vicinity of J092521; this together with the fact that the {\sl WISE} 22\,{\micron} and the {\sl IRAS} 25\,{\micron} fluxes are in good agreement (see Fig.~\ref{fig:diskmodels}), indicates that the {\sl IRAS} detections are also 
likely associated with J092521. Looking through the {\sl AKARI} IRC \citep{ishihara2010} all-sky infrared catalogues, we found that J204315 was measured at 9\,{\micron}.

\begin{table}                                                                  
\begin{center} 
\caption{VISIR observations \label{tab:visirobs}}
\begin{tabular}{lccc}                                                     
\hline\hline
Name      & Obs. date    & Band    & Flux density (mJy) \\
\hline
TYC\,5940 & 2022-01-29 &  B8.7   &  38.3$\pm$2.1     \\
          & 2022-01-29 &  B10.7  &  98.6$\pm$5.0     \\
          & 2022-01-29 &  B12.4  &  30.0$\pm$2.5     \\
\hline				      		   
J060917   & 2022-01-30 &  B8.7   &  20.3$\pm$1.3     \\
          & 2022-01-30 &  B10.7  &  28.6$\pm$1.6     \\
          & 2022-01-04 &  B12.4  &  20.5$\pm$1.5     \\
\hline				      		   
TYC\,8105 & 2022-01-30 &  B8.7   &  57.1$\pm$3.0     \\
          & 2022-01-30 &  B10.7  &  89.8$\pm$4.6     \\
          & 2022-01-30 &  B12.4  &  64.7$\pm$3.8     \\
\hline	  			      		   
J071206   & 2022-01-30 &  B8.7   &  14.7$\pm$1.0     \\
          & 2022-01-30 &  B10.7  &  46.6$\pm$2.5     \\
          & 2022-01-30 &  B12.4  &  15.0$\pm$1.4     \\
\hline				      		   
J092521   & 2022-01-29 &  B8.7   &  25.8$\pm$1.6     \\
          & 2022-01-29 &  B10.7  &  24.4$\pm$1.6     \\
          & 2022-01-05 &  B12.4  &  14.1$\pm$1.4     \\
\hline				      		   
J104416   & 2022-01-31 &  B8.7   &  21.6$\pm$1.4     \\
          & 2022-01-31 &  B10.7  &  45.6$\pm$2.5     \\
          & 2022-02-10 &  B12.4  &  18.3$\pm$1.5     \\
\hline				      		   
TYC\,4946 & 2022-01-31 &  B8.7   &  31.2$\pm$1.8     \\
          & 2022-01-31 &  B10.7  &  28.7$\pm$1.7     \\
          & 2022-02-14 &  B12.4  &  15.3$\pm$1.3     \\   
\hline	  			      		   
J204315   & 2022-05-31 &  B8.7   &  57.6$\pm$3.1     \\
          & 2022-05-31 &  B10.7  &  104.5$\pm$5.4    \\
          & 2022-05-31 &  B12.4  &  38.5$\pm$3.1     \\
\hline
\\
\end{tabular}
\end{center}
\end{table}

\subsubsection{Modelling of the infrared excess} \label{sec:excessmodel}
Fig.~\ref{fig:diskmodels} shows that all five objects exhibit strong IR excess implying the presence of warm 
circumstellar dust. Previous measurements of EDDs have shown that they tend to display substantial variability at 
least at 3--5\,{\micron} \citep{su2019,moor2021,moor2022,rieke2021}. Therefore, initally only the simultaneously measured {\sl WISE} data 
points were used to characterize the excess that are fitted by a single-temperature blackbody model. In the fitting 
we employed Levenberg-Marquardt algorithm using the IDL code developed by \citet{markwardt2009} to estimate the 
characteristic dust temperature and the solid angle of the disc as free parameters. Utilizing the results of an initial fit 
we performed the colour correction of the data points, and then we used the corrected data in the final fitting 
\citep[for more details see,][]{moor2021}. 
This simple model provides good results for J071206, J104416, and J204315 (Fig.~\ref{fig:diskmodels}), yielding 
dust temperatures ($T_\mathrm{d}$) of 434$\pm$25, 405$\pm$21, and 404$\pm$15\,K and fractional luminosities ($f_\mathrm{d} = L_\mathrm{disc}/L_*$) of 0.026, 0.025, and 0.048, respectively. On the other hand, poor quality of the fits for J060917 and J092521 suggested multiple temperature dust 
components in these systems. As Figure~\ref{fig:diskmodels} demonstrates, by adopting a two-temperature model and including 
the 2MASS data in the fit, we have achieved a much more convincing result for J060917, where hot and warm dust components have temperatures of 752$\pm$48 and 237$\pm$22\,K, and fractional luminosities of  0.075 and 0.027. In line with this, a comparison between the values of the corrected Akaike Information Criteria \citep[AIC$_\mathrm{c}$;][]{burnham2002} calculated for the one- and two-component models confirmed the better quality of the latter approach.
The total fractional luminosity of this 
disc is thus $\sim$0.1, which is remarkably high even among EDDs (see Sect.~\ref{sec:agedistribution}). For J092521, \citet{higashio2022} has already proposed a two-temperature disc model based on {\sl WISE} data only. If the {\sl IRAS} data points are also included in the fit, the 
necessity of a second colder dust component is even more striking, and this is also confirmed by our comparison based 
on AIC$_\mathrm{c}$. Using such a model, we obtained $T_\mathrm{d,warm} = 857\pm$90\,K and $f_\mathrm{d,warm} = 0.021$ for the warm and $T_\mathrm{d,cold} = 99\pm6$\,K and $f_\mathrm{d,cold} = 0.033$ for the cold dust. 
{\sl WISE} photometry in all 4 bands are available for this source in a second epoch, half a year after the first. Repeating of the fitting process using this second data set resulted in the following disc parameters: 
$T_\mathrm{d,warm} = 859\pm$69\,K, $f_\mathrm{d,warm} = 0.031$, $T_\mathrm{d,cold} = 97\pm$6\,K, $f_\mathrm{d,cold} = 0.032$.
Our results in both epochs are broadly consistent with the ones obtained by \citet{higashio2022} 
($T_\mathrm{d,warm} = 1108^{+278}_{-393}$\,K, $f_\mathrm{d,warm} = 0.021$, $T_\mathrm{d,cold} = 150^{+37}_{-13}$\,K, $f_\mathrm{d,cold} = 0.016$).

Assuming the presence of blackbody particles, the $R_\mathrm{BB} (\mathrm{au}) = (278.3/T_\mathrm{d} (K))^2 L_* (L_\odot)^{0.5}$ formula \citep{wyatt2008} gives an estimate of the distance of the emitting warm dust from the star. For J071206, where the presence of a nearby companion is very likely, we have assumed that the disc surrounds the primary component. The luminosity of the star was estimated from its effective temperature by intepolation in the table of basic parameters for dwarf stars compiled by \citet{pecaut2013}. For J061709 and J092521, the excess can be described by two-temperature components. In the calculation of $R_\mathrm{BB}$, it was assumed that this indicates the presence of two distinct dust rings, which is, however, not necessarily the case. The warm rings in these systems have $R_\mathrm{BB}$ between 0.05 and 0.3\,au, while in two systems there are also colder dust material at 1--3.7\,au.

The obtained disc parameters are summarized in Table~\ref{tab:props}.

\begin{figure*}
\centering
\includegraphics[width=0.96\textwidth]{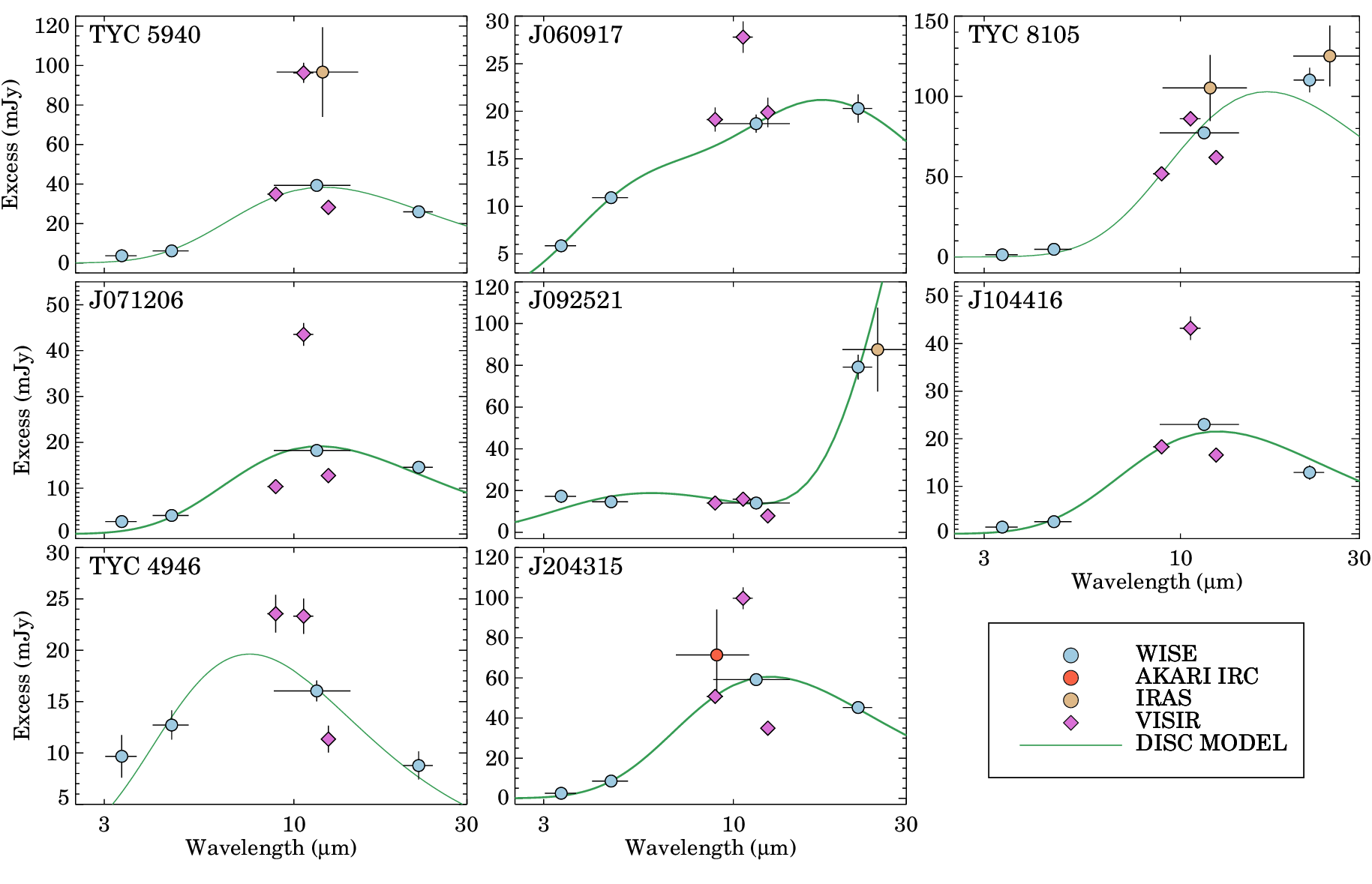}
\caption{Spectral energy distribution of the excess emission. Our new VISIR data points are marked by 
violet diamonds, while the symbols for {\sl IRAS, WISE} and {\sl AKARI} observations are the same as in Fig.~\ref{fig:diskmodels}. Horizontal bars show the width of the filter used. 
The disc models, which are taken from Sect.~\ref{sec:excessmodel} and from \citet{moor2021}, are indicated by green lines.}
\label{fig:excesssed}
\end{figure*}

\subsubsection{Infrared variability} \label{sec:irvar}
To assess whether any of the four newly discovered EDDs or J092521 exhibited significant variability at 3.4 and/or 4.6\,{\micron} 
between 2013 and 2022, we analysed their single-exposure $W1$ and $W2$ band photometric measurements 
downloaded from NEOWISE-R Single Exposure Source Table archived by IRSA. After we removed data points of dubious quality using 
the flags assigned to the measurements and following the method proposed by \citet{moor2021}, we computed the $\chi_\mathrm{red}^2$ tests 
for each band individually and the Stetson $J$ ($S_J$) index that measures correlated variability in the $W1$ and $W2$ bands 
\citep{sokolovsky2017}. To determine whether the discriminants obtained in this way indicate significant changes, it is necessary 
to define a threshold, which is derived in each case by comparison with the ensemble of variability indices calculated for {\sl WISE} sources of 
similar brightness ($\Delta W_1 = \pm1$\,mag, $\Delta W_2 = \pm1$\,mag) located within 1\fdg5 of the target.
After identifying and removing outliers using a sigma clipping procedure, the means ($D_{\chi_\mathrm{red,W1}^2}$, 
$D_{\chi_\mathrm{red,W2}^2}$, $D_{S_J}$) and the standard deviations ($\sigma D_{\chi_\mathrm{red,W1}^2}$, 
$\sigma D_{\chi_\mathrm{red,W2}^2}$, $\sigma D_{S_J}$) were 
calculated for the comparison sample, and the threshold was calculated as $D_x + 5 \sigma D_x$ \citep[see also in][]{moor2021}.

We concluded that all of our targets are variable. J060917 and J092521 show significant and correlated changes in both bands, 
while for the other three objects, the variability is limited to the $W2$ band.
In the $W1$ and $W2$ bands both the star and the circumstellar disc contribute to the observed emission. With the exception of 
J060917, where the contribution of the disc is $\sim$48\% and $\sim$76\% of the total flux density at 3.4 and 4.6\,{\micron}, respectively, 
for the other objects the stellar photosphere dominates the emission in both bands.  
It is therefore worth scrutinizing whether possible changes of the 
host stars could explain the variability observed in the {\sl WISE} light curves. By examining the available ASAS-SN light curves, we found that the brightness of the targets was stable over the given timescale and did not show any similar pattern to the changes seen in the {\sl WISE} 
data streams. That the observed variations are related to the discs is also supported by the fact that their amplitudes, in correlation with the larger contribution of the disc to the total flux, are always larger 
in the W2 than in the W1 band (or can only be detected in the W2 band). 

By averaging the individual single exposure data points in each {\sl WISE} observing window between 2013 and 2022 and by subtracting the   
stellar contribution from the measured flux densities using the photosphere models (Sect.~\ref{sec:photmodels}), we compiled and plotted 
the $W1$ and $W2$ light curves of the five discs that are found to be variable (Fig.~\ref{fig:irvar}).

\section{VLT/VISIR observations} \label{sec:visirobs}
In order to assess whether the 8 targeted EDDs display N-band solid state features, we used the 
VLT Imager and Spectrometer for the mid-InfraRed \citep[VISIR,][]{lagage2004} on ESO's Very Large
Telescope (VLT, UT2) to measure their flux densities in three bands at 8.91\,{\micron} (B8.7), 10.64\,{\micron} (B10.7), and 12.45\,{\micron} (B12.4). The observations were carried out under 
ESO programme 106.212F.001 (PI: A. Mo\'or). We used the regular imaging mode with the Small
Field objective that provides a field of view of 38\arcsec$\times$38\arcsec with a pixel scale of 45\,mas$\times$45\,mas\footnote{\url{https://www.eso.org/sci/facilities/paranal/instruments/visir/doc/VLT-MAN-ESO-14300-3514_2020-03-03.pdf}}. To subtract the sky background, chopping and nodding was used with a chop throw of 8{\arcsec} 
perpendicular to the nodding direction. While for the brightest sources all three band measurements were taken at the same time in the same run, 
fainter objects were observed in two separate runs (see Table~\ref{tab:visirobs}). 
Using the same filters, immediately before or after our main targets' measurements, 
we also observed a nearby mid-IR standard star that was selected from the catalogue 
compiled by \citet{cohen1999}. The average airmass difference between the target and the corresponding 
standard object was 0.03, with a maximum difference of 0.15.

We used the EsoReflex GUI environment \citep{freudling2013} with VISIR 4.4.2 pipeline kit\footnote{\url{https://ftp.eso.org/pub/dfs/pipelines/instruments/visir/visir-reflex-tutorial-1.5.pdf}} 
and the \texttt{visir\_img.xml} workflow to reduce 
and calibrate our raw data sets. The pipeline produced a flux calibrated image as well as an 
uncertainty image. All our targets in all bands have been detected as single point sources in the images. 
In most cases the objects have circularly symmetric brightness profile, the only exception is J204315, which due to its 
position in the sky, has been measured at relatively large zenith distances and therefore it has an elongated shape 
especially at 8.91 and 10.64\,{\micron}. The average full width at half-maximum of the sources is $\sim$6, $\sim$7, and $\sim$8\,pixels (272, 317, and 362\,mas) in 8.91, 10.64, and 12.45\,{\micron}. To extract their flux densities we applied aperture photometry 
with a radius of 6\,pixels (10\,pixels for J204315) in all bands. The background was estimated in an annulus between 
20 and 30\,pixels. To determine the internal photometric error we used the uncertainty image produced by the pipeline. 
The aperture correction factor for a given observation was derived using the associated standard star measurement. 
By comparing the flux densities obtained for our standard stars with the expected values taken from the literature
\citep[][and the table available on the VISIR homepage\footnote{https://www.eso.org/sci/facilities/paranal/instruments/visir/tools.html}]{cohen1999}, we found that they reproduce them with an $\sim$5\% average accuracy. We therefore calculated the final photometric uncertainty as the quadratic sum of the internal measurement error and this calibration uncertainty. The obtained flux densities and their uncertainties are listed in Table~\ref{tab:visirobs}.

\section{Results} \label{sec:results}
In Figure~\ref{fig:excesssed} we present the SEDs of the excess emission for all 8 targets. In addition to the available broad-band photometry, 
the narrow-band VISIR data are also plotted (the horizontal bars show the width of the used photometric bands\footnote{The effective widths of the 
filters are taken from the SVO filter profile service (\url{http://svo2.cab.inta-csic.es/theory/fps/}).}.
For TYC~5940, TYC~8105, and 
TYC~1213, mid-IR data other than the VISIR photometry and the photospheric models used to compute the excesses are taken from \citet{moor2021}. As can be seen, VISIR data points typically peak in the B10.7 band and show abrupt changes in the narrow spectral region examined, indicating the presence of solid state features in the SED.

\begin{figure}
\centering
\includegraphics[width=0.49\textwidth]{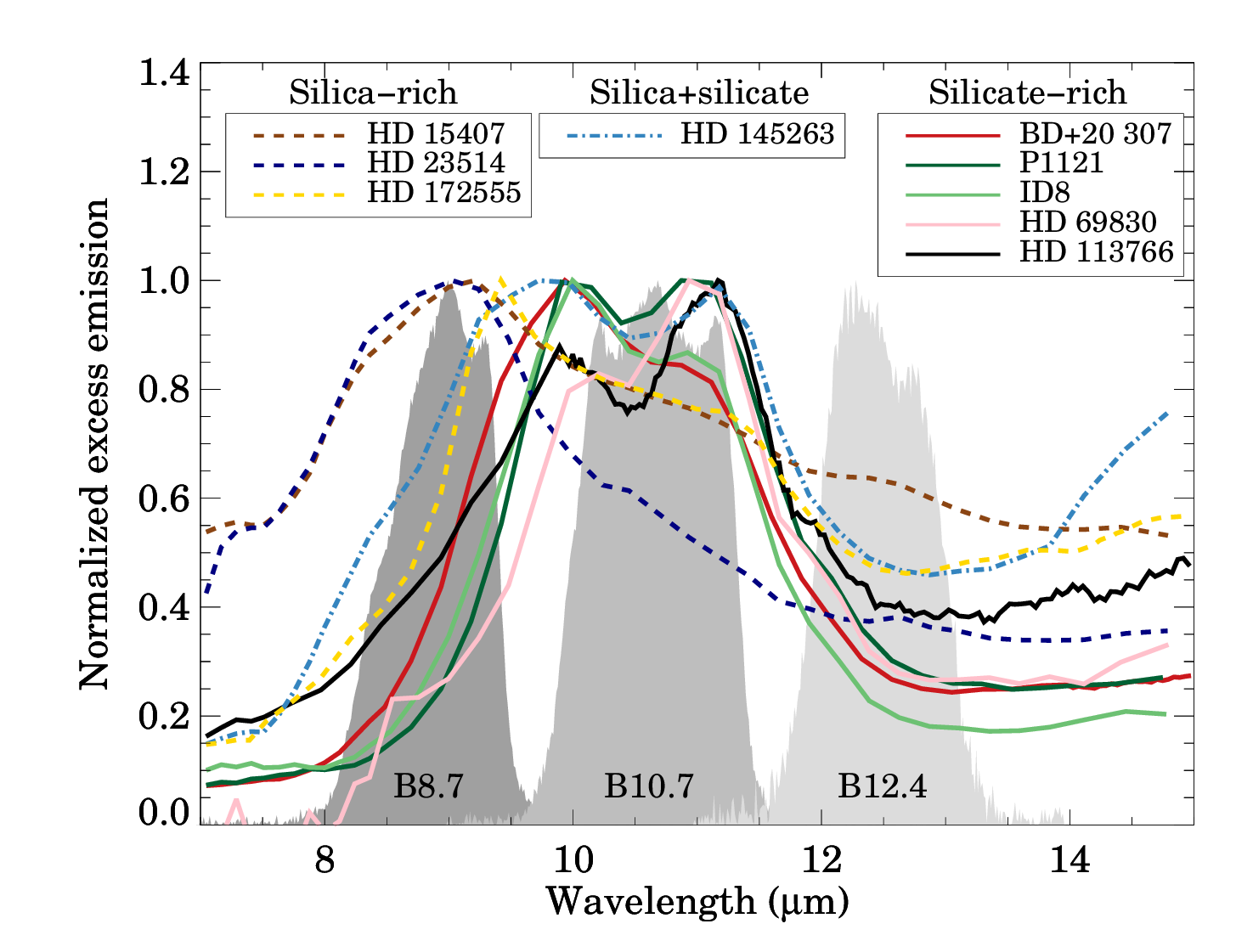}
\includegraphics[width=0.49\textwidth]{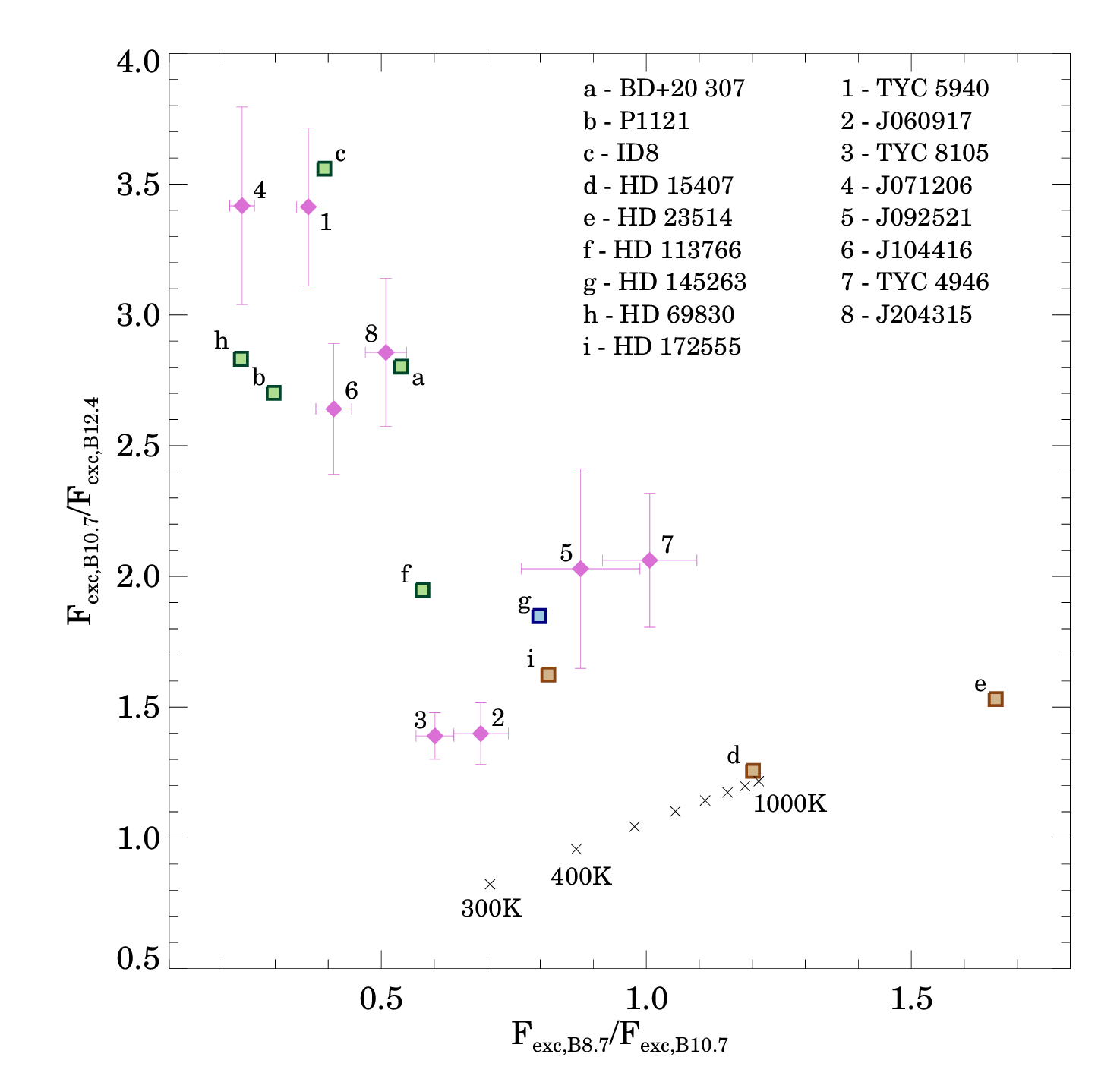}
\caption{Top: Normalized excess spectra between 7 and 15\,{\micron} of nine debris discs with transient origin. 
{The original {\sl Spitzer} spectra, from which they were derived, were taken from the literature \citep[][see Sect.~\ref{sec:results} for more details]{chen2006,lebouteiller2011,su2020}.}
Transmission curves of the three applied band-pass filters (B8.7, B10.7, and B12.4) are also displayed.
Bottom: $F_\mathrm{exc,B10.7}/F_\mathrm{exc,B12.4}$ flux ratios as a function of 
$F_\mathrm{exc,B8.7}/F_\mathrm{exc,B10.7}$ ratios for our eight targets (purple diamonds) 
and for the nine abovementioned discs (squares) with high quality mid-IR spectra obtained
with the {\sl Spitzer}/IRS spectrograph. For the latter objects, synthetic photometry was used to 
derive the flux ratios, the silicate- and silica-dominated discs are marked in green and brown, respectively, 
while HD\,145263 is marked in blue.} 
\label{fig:fluxratios}
\end{figure}

To investigate this further, we compared our results with the outcomes of mid-IR spectroscopic studies carried 
out by {\sl Spitzer} for nine warm debris discs showing prominent solid-state features (Fig.~\ref{fig:fluxratios}, top). 
Seven objects (BD+20~307, P~1121, ID~8, HD~113766, HD~15407, HD\,23514, and HD\,145263) from the selected sample are classified as EDDs in 
the literature. Although the fractional luminosity of the other two discs, around HD\,69830 and HD\,172555, is substantially lower 
\citep[2.4$\times$10$^{-4}$ and 7.2$\times$10$^{-4}$, respectively,][]{mittal2015} than that of EDDs, their warm dust material is thought to have transient origin \citep{beichman2005,wyatt2007,lisse2009,johnson2012}.

Looking at the spectral range we have examined, between 8 and 13\,{\micron}, two groups can be broadly distinguished on the basis of the observed solid state features (Sect.~\ref{sec:intro}). Five of these discs (BD+20~307, P~1121, ID~8, HD~113766, and HD\,69830) have strong solid-state emission features between 10 and 11.5\,{\micron}, implying the presence of fine amorphous and crystalline silicate dust particles \citep{olofsson2012,morlok2014,meng2015}. The spectra of HD~15407 and HD\,23514 peak at $\sim$9\,{\micron}, indicative of copious amount of small silica grains \citep{rhee2008,olofsson2012,morlok2014,meng2015}. The disc around HD\,172555 is also rich in silica as evidenced by a dominant feature at $\sim$9.3\,{\micron} \citep{lisse2009}.  
The HD\,145263 represents a kind of transition between the two groups with its mid-IR spectrum 
indicate the presence of some silica in the disc in addition to silicates \citep{lisse2020}.
For comparison with our data, we had to extract synthetic photometry from the available low-resolution IRS spectra. With the exception of HD~113766 and 
HD\,172555, the spectra were retrieved from the LR7 release of the Cornell AtlaS of Spitzer/IRS Sources \citep[CASSIS;][]{lebouteiller2011} database, which provides uniformly processed, high-quality IRS spectra. Being spatially unresolved, for all seven sources, we selected the data product obtained using the optimal extraction. 
HD~145623 was observed three times with the IRS. Since these measurements show no variability around 10\,{\micron}, we finally chose 
the middle observation (AOR: 21808896) for our analysis. 
HD\,69830, which was the subject of a multi-epoch mid-IR spectroscopic monitoring 
project carried out by the {\sl Spitzer} \citep{beichman2011}, 
has 6 low-resolution IRS spectra in the CASSIS database, from which we used 
the one with the AORkey of 28830720. 
In the case of HD\,113766 and HD\,172555, where the IRS observations were performed in spectral mapping mode, and thus is not included in CASSIS, the spectra were taken from the literature \citep{chen2006,su2020}. 
From these spectra then, we computed synthetic photometry in the three VISIR bands using the relevant filter profiles downloaded from the VISIR homepage\footnote{\url{https://www.eso.org/sci/facilities/paranal/instruments/visir/inst.html}}. To calculate excesses, we also needed the stars' photospheric models, which were taken from \citet{moor2021}. IRS spectra between 7 and 15 {\micron} of the nine discs as well as 
the three filter profiles are displayed in the top panel of Fig.~\ref{fig:fluxratios}.

Figure~\ref{fig:fluxratios} (bottom) shows the flux ratios of $F_\mathrm{exc,B10.7}/F_\mathrm{exc,B12.4}$ as a function 
of $F_\mathrm{exc,B8.7}/F_\mathrm{exc,B10.7}$ ratios for our targets (violet diamonds) and for the selected nine discs observed by the 
{\sl Spitzer} (squares, green for silicate-dominated discs, brown for systems with silica features, and blue for HD\,145263). 
The measured flux ratios of our targets (violet diamonds) are substantially different from those of plain blackbodies with temperatures between 300 and 1000\,K (black crosses). TYC~5940, J071206, J104416, and J204315 coincide well with 
four silicate-dominated discs (BD+20~307, P~1121, ID~8, and HD\,69830), indicating that their solid state 
emission features may be broadly similar to those detected in the spectra of the latter objects. 
The $F_\mathrm{exc,B8.7}/F_\mathrm{exc,B10.7}$ flux ratios of the other four discs are higher. They 
occupy roughly a region in the figure where both silicate-rich and silica-rich discs from the {\sl Spitzer} sample are found, 
and thus they cannot be classified with certainty on the basis of our current data.
In the case of J092521 and TYC\,4946, the $F_\mathrm{exc,B8.7}/F_\mathrm{exc,B10.7}$ flux ratios 
exceed that measured for HD\,145263 but is below that of HD\,15407 
and especially HD\,23514. This and the proximity of the two objects to HD\,145263 in the figure 
raises the possibility that these discs also contain some silica grains. However, 
their difference from HD\,23514 suggests that silica, if present at all in them, 
may be much less dominant than in the latter EDD.   

Based on our measurements, thus, the majority of the sources studied may show strong solid-state features around at least at 
10\,{\micron}. Because of this, the true shape of the SED may even differ substantially from the blackbody model fitted to the broadband photometric data, and thus the true dust temperatures may show deviations from the fitted values that could be even significantly larger than the formal errors derived in Sect.~\ref{sec:excessmodel}.

\section{Discussion} \label{sec:discussion}

\subsection{Age distribution and age related trends of EDDs}  \label{sec:agedistribution}
The four new discoveries presented here together with the 20 discs known from the literature (including the other 
four objects we analyse in the paper) raises the total number of known EDDs around F-K-type main-sequence stars to 
24 (see Table~\ref{tab:eddnames}). 
In the following, we review their basic characteristics, their dependence on age and examine the age distribution 
itself. 

\begin{figure}
\centering
\includegraphics[width=0.49\textwidth]{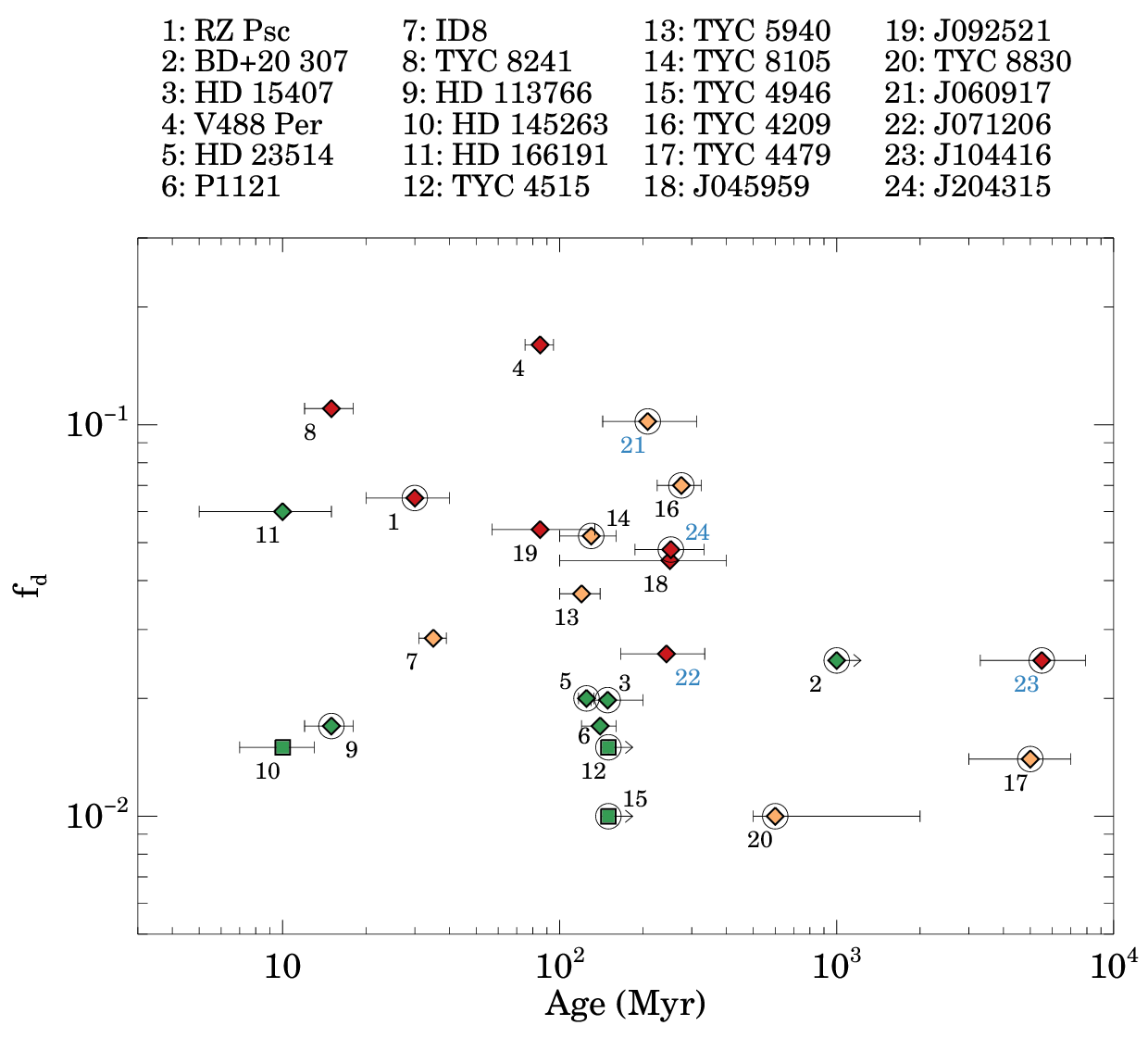}
\caption{Fractional luminosity as a function of age for 20 EDDs from the literature (marked by black numbers) and for our four new discoveries (marked by blue numbers).
Spectral type of the host stars are indicated by colours (F-type: green, G-type: light orange, K-type:red). 
Objects with significant changes at 3--5\,{\micron} are marked with diamond symbols while non-variables are marked with 
squares. Multiple systems are indicated by larger circles around the symbols. Due to their variability, the $f_\mathrm{d}$ of the discs is likely to vary, but the extent of this is difficult to estimate, as our time-domain data are mostly limited to a narrow wavelength interval between 3 and 5\,{\micron} range. The TYC\,8241 disc has depleted significantly after 2010 \citep{melis2012}, the data point shown corresponds to the pre-2010 state.
Some of the object identifiers above the figure are abbreviated; therefore, in Table~\ref{tab:eddnames}, we provide 
the corresponding AllWISE identifiers of all systems and, where available, their SIMBAD-compatible names.} 
\label{fig:fdage}
\end{figure}

Figure~\ref{fig:fdage} shows the fractional luminosity of EDDs as a function of age. The data for this plot were taken 
from Table~\ref{tab:props} and from the literature \citep[][and references therein]{vican2016,tajiri2020,melis2021,moor2021}. 
EDDs are typically defined as objects with $f_\mathrm{d} \gtrsim 0.01$, and thus they represent the highest fractional 
luminosity tail of the transient debris disc population. An observable feature of the distribution is that the $f_\mathrm{d}$ 
of discs hosted by F-type stars (indicated by green symbols) tend to be smaller than those around G- (light orange) and 
K-type (red) stars. This does not necessarily mean there is less dust in EDDs around the F-type 
stars.  This trend may be partly related to the stronger radiation pressure of the more luminous F stars, which can thus blow out 
much larger grains than in the case of G, K stars, thereby resulting in a smaller $f_\mathrm{d}$ for the same dust composition, 
mass and radial distance. Another explanation could be that the average radial distance of the EDDs around F-type stars is larger 
than around later spectral type hosts. The available $R_\mathrm{BB}$ data hint at such a trend, but as the uncertainty 
may be particularly high for systems with sparsely sampled SEDs, no firm conclusions can be drawn at present.
The high fractional luminosity of the EDDs and the complex mid-infrared light curves they exhibit suggest that there 
are regions in the discs that are optically thick \citep{kennedy2014,meng2014,su2019} at least at wavelengths shorter than a 
few microns, which may further complicate this issue.
Another notable trend in Figure~\ref{fig:fdage} is that the majority of discs around G- and K-type stars younger 
than 300\,Myr have  $f_\mathrm{d} > 0.03$, while older EDDs with similar host stars (although very few are known) 
exhibit $f_\mathrm{d}$ lower less than this value.

Thanks to the NEOWISE reactivation mission, light curves with a cadence of a half year at 3.4 and 4.6\,{\micron} 
are available for all EDDs, allowing the investigation of possible annual variations in the last decade. In addition, some EDDs were also monitored with {\sl Spitzer} at 3.6 and 4.5\,{\micron} 
\citep[e.g.][]{meng2014,su2019,su2020,su2022}. By studying the 17 EDDs known at that time, \citet{moor2021} found 
that 14 of them show significant variability in the 3--5\,{\micron} range, i.e. this is an inherent property of EDDs.
The observed rapid changes of EDDs are mostly attributed to the dynamical and collisional evolution of a cloud of fresh dust 
produced after vaporisation and subsequent condensation of material ejected in a high velocity 
impact \citep{jackson2014,su2019,su2022,watt2021,moor2022,lewis2023}.
The NEOWISE light curves of five of the seven objects discovered since then have been analysed in 
Sect.~\ref{sec:irvar}. For the other two EDDs, TIC\,43488669 \citep[J045959;][]{tajiri2020} and TYC\,8830-410-1 
\citep[TYC\,8830;][]{melis2021}, proceeding in a similar spirit, we found that both discs are variable 
(see Fig.~\ref{fig:irvar} for light curves of the discs). In Figure~\ref{fig:fdage}, variable objects -- 
the majority of the sample, 21 systems out of the 24 -- are marked by diamonds, while those displaying no 
significant changes are shown by squares. This statistic further reinforces the abovementioned finding that 
mid-IR variability is a fundamental characteristic of EDDs.

Current theories mostly link the formation of EDDs to giant collisions taking place in the region of terrestrial 
planets \citep{melis2016,su2019}. Such collisions between large planetary embryos are thought to be common during 
the final chaotic growth phase of rocky planets, raising the possibility that EDDs may indeed be the result of these events.
However, according to simulations, these planet-forming processes occur mostly in the first tens of 
million years, and the frequency of collisions drops off significantly after 100\,Myr \citep[e.g.][]{quintana2016}. 
Consistent with this picture, recent studies suggest that the era of giant impacts in the Solar System probably ended by 80\,Myr \citep{woo2022}. 
Looking at the age distribution of the currently known EDDs (Fig.~\ref{fig:fdage}), we see that their incidence tends to 
decrease very slowly in the first 300\,Myr:
there are 8 objects younger than 100\,Myr, 7 with ages between 100 and 200\,Myr 
(5 if we assume that the two objects with only a lower age limit of 150\,Myr are indeed much older), and 5 
between 200 and 300\,Myr. Considering not only perfect accretion but fragmentation and hit-and-run collisions in their 
simulations of rocky planet formation around Sun-like stars, \citet{quintana2016} found (see their fig.~5) that 
$\sim$95\% of the large collisions happen in the first 100\,Myr and $\sim$4\% between 100 and 200\,Myr. 
This is very different from what we see in the observed sample and suggests that the formation of EDDs is not exclusively related to the formation of rocky planets, but other mechanisms, some of which are activated later, may also play an important role. 
\citet{hansen2023} raises the possibility that the tidal evolution of a planet-moon system can lead to the escaping of the moon, which can then collide with its original parent planet. Such process could result in collisions on a similar scale to those that are thought to occur during formation of the rocky planets, but at a much later time, thereby explaining the existence of older EDDs.
Another alternative explanation is that the increased dust production may be linked to some sort of rearrangement of the planetary system. 
This can also occur in young systems, for example, the dusty disc of RZ Psc is proposed to be related to the migration of recently formed 
giant planets, that might be initiated by the known companion (see below) of the star \citep{su2023}. 
In an older system, a late dynamical instability can trigger processes that
cause significant episodic dust production in the terrestrial zone via large collisions, and/or sublimation of a swarm of icy bodies 
transported into that region from an outer reservoir \citep[e.g.,][]{melis2021,moor2021}. 

Previous studies suggested that wide-separation companion stars may play a 
role in the formation of warm dust-rich discs \citep{zuckerman2015,silverberg2018,moor2021}.
In Figure~\ref{fig:fdage}, known multiple systems are indicated by larger circles around the symbols.
For TYC\,5940, J092521, and in particular J071206, the Gaia\,DR3 radial velocity or astrometric data strongly suggest the 
presence of companion stars (Sect.~\ref{sec:multiplicity}), but these are not yet considered secure results and are therefore 
not marked in the figure.
Two of the eight EDD host stars younger than 100\,Myr are known to be binaries, RZ\,Psc has a companion at a projected separation 
of $\sim$20\,au \citep{kennedy2020}, while the projected separation between the two components of the HD\,113766 system, based on their Gaia DR3 data, is $\sim$150\,au. At least 75\% (12/16) of objects older than 100\,Myr have a companion, their projected separations 
range between 365 and 10261\,au\footnote{BD+20~307 is a triple system consisting of a close inner binary 
\citep{weinberger2008,zuckerman2008} and a distant tertiary component with a projected separation of $\sim$980\,au, 
as recently revealed using Gaia~DR2 astrometric data \citep{moor2021}.}.   
In this older subgroup, thus, the fraction of binaries with separation $>$300\,au is significantly higher than among the 
younger EDDs. This incidence of wide binary pairs is also significantly higher than that observed for normal main sequence stars with similar spectral type \citep{moor2021}. Remarkably, all EDD host stars older than 500\,Myr have known wide 
companions.

\begin{figure}
\centering
\includegraphics[width=0.49\textwidth]{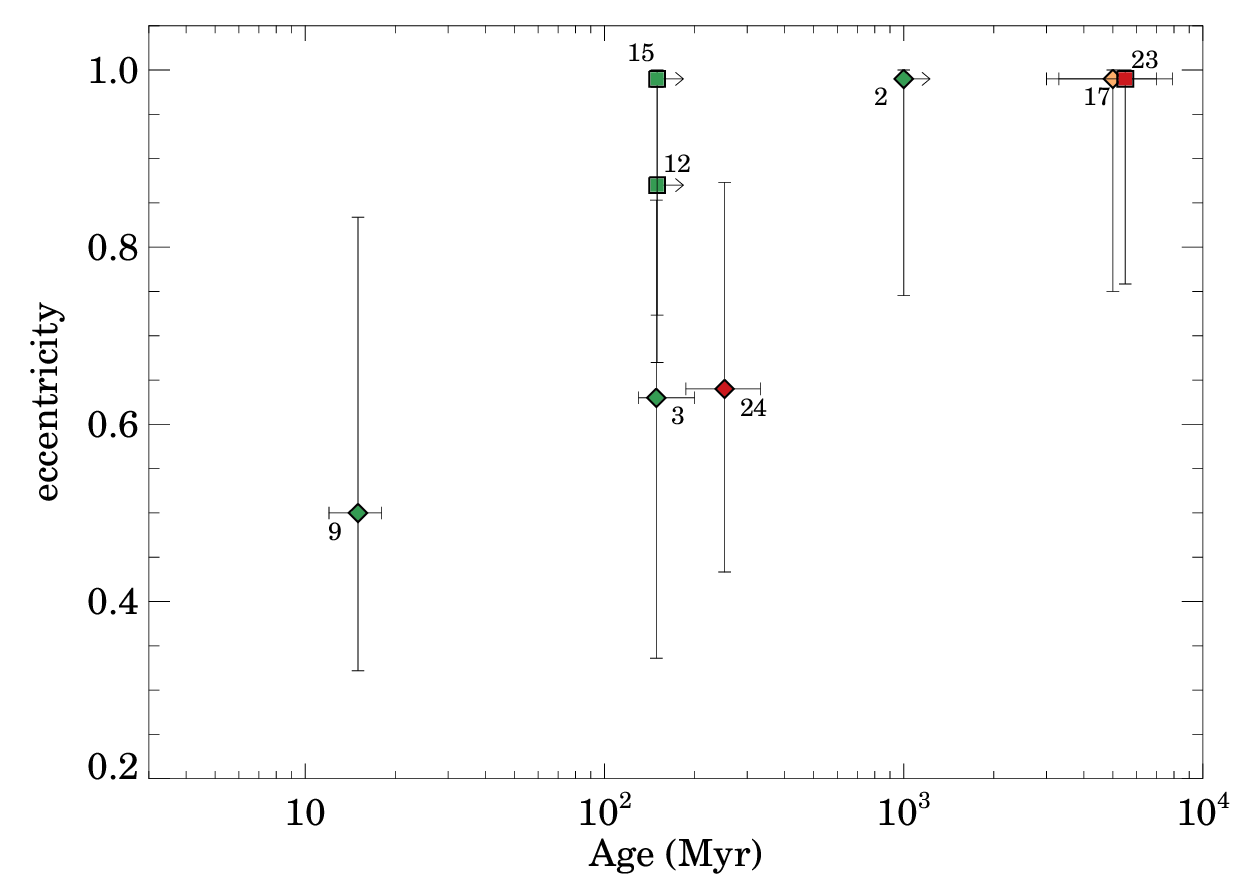}
\caption{Estimates of eccentricities of 10 binary EDD systems based on \citet{hwang2022a}.
The vertical error bars correspond to the lower and upper eccentricity limits of the 68\% 
credible interval. See Fig.~\ref{fig:fdage} to identify which objects the numbers represent.} 
\label{fig:eccen}
\end{figure}

This observed characteristic of older EDDs suggests that their wide companions might have played a role 
in the formation of the disc, for example by triggering a late dynamic instability in their planetary systems. 
Although the companions are currently very far away from the discs, wide binary systems, in particular the equal-mass pairs, tend to 
have highly eccentric orbit \citep{tokovinin2020,hwang2022a,hwang2022b}, i.e. it is possible that the processes that led to 
the disc's formation were triggered during the last pericentre passage of the pair. A recent catalogue compiled by \citet{hwang2022a} using Gaia astrometry contains estimates of the eccentricities of 10 binary EDD systems.
Figure~\ref{fig:eccen} shows the most probable eccentricities and the lower and upper eccentricity limits of the 68\% 
credible interval for those 8 object, where all quality criteria proposed by \citet{hwang2022a} (in Sect.~4.3 of their work) 
are fulfilled. For the three oldest EDDs with ages $\gtrsim$1\,Gyr (BD+20 307, TYC\,4479, J104416) and for TYC\,4946, which has only a lower age estimate of 150\,Myr, the best eccentricity estimate is 0.99, with a confidence interval upper bound of 1 and lower bounds 
of $\sim$0.75. With such large eccentricities, even with the relatively large projected separations of 10$^3$--10$^4$\,au 
measured for these systems, it is plausible to assume that the (probably most recent) pericentric passage of the companion could have significantly perturbed the possible planetary systems.
Thus, these objects could be promising candidates that are just now undergoing a dynamic instability process, e.g. akin to the hypothesized Late Heavy Bombardment in the Solar system.   

The orbital plane of the companion star may also be different from that of the planetary system including the planetesimal disc(s).
In case of sufficiently high mutual inclination, the companion can affect the orbit of the smaller bodies via the 
Kozai-Lidov mechanism, leading to oscillation in their eccentricity and inclination \citep{naoz2016}. 
This could cause stronger excitation of the disc, more frequent collisions and hence enhanced dust production \citep{nesvold2016,moore2023}, or even drive icy planets in the outer belt into the inner regions by increasing 
their eccentricity, where they can contribute to the dust disc via their sublimation and disruption \citep[e.g.][]{young2023}. 
Whether these mechanisms can indeed explain the very high dust content of EDDs require further investigations.

\subsection{Solid-state feature from sub-micron grains}
In addition to the {\sl Spitzer} IRS observations mentioned in Sect.~\ref{sec:results}, ground-based mid-IR spectroscopic observations around 
10\,{\micron} are available for four EDDs: RZ\,Psc \citep{kennedy2017}, V488\,Per \citep{sankar2021}, HD\,166191 \citep{kennedy2014}, and 
TYC~8830-410-1 \citep{melis2021}. Of these, only V488\,Per shows no detectable solid-state emission, the other three 
 display clear solid-state features peaking between 10 and 11\,{\micron} indicating the presence of small amorphous and crystalline 
 silicates in their discs \citep{kennedy2014,kennedy2017,melis2021}. 
 By taking into account the outcome of our multiband photometric 
 programme as well,
 now we have information on solid-state emission around 10\,{\micron} for 19 of the 24 EDDs identified so far. 
 Even in this larger sample, 
 V488\,Per is the only one that is probably featureless. 
For the vast majority of EDDs' mid-IR spectra, the most dominant solid-state features present between 10 and 11.5\,{\micron}, 
indicating silicate-rich dust material, with only two systems (HD\,15407, HD\,23514) so far known where the spectra are 
dominantly characterised by features between 9 and 9.5\,{\micron} as an indication of significant amount of silica grains. 
In addition to them, HD\,145263 exhibits both silica and silicate emission features \citep{lisse2020}, and based on our 
VISIR multiband observations, we have identified two other EDDs, J092521 and TYC\,4946, that may have a similar  
characteristic.
By comparing mid-IR {\sl Spitzer} spectra of warm debris discs with the spectra of solar system rocky samples, 
\citet{morlok2014} argued that the material of silicate-dominated discs are more mantle-type while the silica-bearing 
objects are associated with crustal-type materials. 

An interesting question is whether the above groups based on solid-state emission features show any correlation with the mid-IR time-domain behaviour of the objects \citep[see also in][]{meng2015}. 
The level of the variability can be characterised in many ways. In the following we generally use the $\chi^2$ statistic of the 
$W2$ NEOWISE Reactivation light curves of the sources as a metric for this purpose.
Among the silica-bearing EDDs, the {\sl WISE} light curves of HD\,15407 show no variability, although due 
to its high brightness the measurements are saturated which degrades the quality. However, based on its {\sl Spitzer} 
time-domain data, a weak but significant variability was detected \citep{meng2015}. 
HD\,23514 shows moderate flux level changes, while for HD\,145263, neither its {\sl WISE} nor its {\sl Spitzer} data show 
significant variations. Out of the two newly identified silica-bearing candidates, TYC\,4946 is also non-variable, while 
J092521 is one of the five most highly variable objects (along with RZ\,Psc, ID\,8, V488~Per, and TYC\,4209). Overall, it seems that the mid-IR variability level of the objects and candidates belonging to this group seems to be rather low. The light curves of the silicate-rich EDDs -- here we include those four objects from the VISIR mini-survey for which there is a strong indication to have this property (TYC~5940, J071206, J104416, and J204315, see Sect.~\ref{sec:results}) -- all show variations, and ID8 and RZ\,Psc are among the objects that have particularly high variability. The average level of variability in this group is higher than for silica-bearing discs \citep[see also in][]{meng2015}.
Finally, V488\,Per, with the most extreme amplitude and very rapid changes, was found by \citet{sankar2021} to be featureless based on its 8--13\,{\micron} spectrum. Considering this result they suggest the presence of dust composed primarily of metallic iron in this disc.

Silica grains are thought to be produced via condensation from silica-rich vapour formed in high-velocity (5--10~km~s$^{-1}$)
giant impacts \citep[e.g.,][]{lisse2009,lisse2012} or in hypervelocity ($\gtrsim 10$~km~s$^{-1}$) collisions between sub-millimeter-sized particles \citep{johnson2012}. Based on these hypotheses, one would expect some correlation 
between the presence of silica and variability since the latter phenomenon is also interpreted as an aftermath of high-velocity 
collisions \citep{meng2015,su2019}. Contrary to this expectation, however, silica-bearing discs typically show only weak mid-IR changes.
A possible explanation for this is that silica particles can also be formed in other ways.
\citet{lisse2020} argued that in the disc of HD\,145263, space weathering triggered by a stellar super-flare 
altered the original dust material, leading to the appearance of such particles. However, for the silica material of HD\,23514 and HD\,15407, 
they also favour a collisional origin. 
For the latter discs, the IRS observations were performed in 2008, while most
{\sl Spitzer} or {\sl WISE} photometric data at 3--5{\micron} were obtained after 2010. Models of post-impact processes suggest that the most pronounced brightness changes in the resulting debris disc occur within a few orbital periods after the impact, which for most EDDs takes a few years at most. If the collisional evolution and grinding of the resulting silica dust population takes longer than this, then it is possible that the impact event and the most substantial brightness changes already took place well before the spectroscopic observations, in a period when no monitoring data were available.

\section{Summary} \label{sec:summary}
Extreme debris discs are peculiarly bright warm circumstellar dust structures around some 
main sequence stars. They have transient origin and likely represent the outcome of a 
giant collisional event that happened within the terrestrial region. 
To further understand this phenomenon, here we present the discovery of four new EDDs. 
By complementing them with four 
recently identified systems from the literature we perform multi-band mid-IR photometry on the whole 
sample with the VLT/VISIR camera to examine whether they exhibit solid-state features around 
10{\micron}. 

Our new findings bring the EDD sample to 24.
For the four new discoveries and one additional object, we provide a detailed analysis of the 
properties of the hosts stars and the 
discs. Based on the age estimates, we find that J104416, one of the newly identified systems, is 
probably the oldest known EDD with its age of $\sim$5.5\,Gyr. 

The obtained photometry at 
8.7, 10.7, and 12.4\,{\micron} implies that all eight targets likely exhibit distinct 10\,{\micron} feature. 
This further strenghtens the previously recognized trend that EDDs tend to display prominent solid-state features 
indicative of abundant sub-micron-sized dust grains. 
Considering our new results, out of the 19 systems where this characteristic could be examined, this is true for 18.
As mid-IR spectroscopic studies of previously known EDDs showed, the majority of them 
have 8--13\,{\micron} spectra charaterized by strong features peaking between 9.9 and 11.3\,{\micron}, 
suggesting dust material dominated by silicate particles, and only few show dominant
feature peaks between 9 and 9.5\,{\micron} indicating the presence of silica grains. 
Based on the measured VISIR flux ratios, we have identified 4 discs in our sample that most probably 
belong to the silicate dominated subgroup. The classification of the other four objects is less certain; 
nevertheless, two of them may contain some silica.

As shown by the available light curves at 3--5\,{\micron}, mid-IR variability on annual time-scales is a common feature of 
EDDs. We concluded that the variability level of discs belonging to the silicate-rich group is typically 
higher than that of silica-bearing systems. \citet{meng2015} found similar results using a smaller sample 
of EDDs. This seems somewhat surprising, given that both the variability and the silica 
formation are thought to be due to the aftermath of high velocity collisions. It is possible, however, that 
silica dust can be formed in other ways \citep[e.g.,][]{lisse2020} and that the lifetime of the 
emerged silica material can be longer than the timescale of the processes leading to the mid-IR variability of the disc.  

According to most current theories, EDDs are probably related to the accumulation of 
terrestrial planets, which is thought to be characterised by giant collisions between large planetary 
embryos. However, the age distribution we obtained for the 24 known systems strongly suggests that the formation 
of EDDs is unlikely to be limited to this process. While numerical simulations of the rocky 
planet formation and experiences in our Solar System imply that the vast majority of such giant impact events 
occur in the first 100\,Myr, the incidence of EDDs begins to decrease only after 300\,Myr, and we know of a few
objects older than 500\,Myr as well. Remarkably, we have found that the majority of EDDs older than 100\,Myr 
-- at least 12 out of the 16 -- are located in wide binary systems. Eccentricity constraints from Gaia astrometry 
are available for 7 of these binaries. For four of them, including the three oldest EDDs, the best eccentricity 
estimate is 0.99. For such eccentric orbits, even with large separations, it is conceivable that the companion 
star could have been able to significantly perturb the planetary system during its pericentre passage,
triggering collisions among bodies located there. Such a late externally induced dynamical instability can provide a possible explanation
for the formation of old EDDs. It is worth noting that in the case of sufficiently high mutual inclination, the companion 
star can perturb substantially the orbital elements of the planetesimals and possible planets via the Kozai-Lidov mechanism, 
which can also trigger processes leading to significant dust production.

Future spectroscopic observations of these discs between 5 and 28\,{\micron} with the 
Mid-Infrared Instrument onboard the {\sl James Webb Space Telescope} would allow their 
more detailed mineralogical analysis providing insights into the origin of their 
warm dust material.

\section*{Acknowledgements} 
We thank the referee for the constructive comments and suggestions 
that greatly improved the paper.
This publication makes use of data products from the Wide-field Infrared Survey 
Explorer, which is a joint project of the University of California, Los Angeles, 
and the Jet Propulsion Laboratory/California Institute of Technology, and NEOWISE, 
which is a project of the JetPropulsion Laboratory/California Institute of Technology. 
WISE and NEOWISE are funded by the National Aeronautics and Space Administration.
The publication makes use of data products from the Two Micron All Sky Survey, 
which is a joint project of the University of Massachusetts and the Infrared
Processing and Analysis Center/California Institute of Technology, funded by the 
National Aeronautics and Space Administration and the National Science Foundation.
This work has made use of data from the European Space Agency (ESA) mission
{\it Gaia} (\url{https://www.cosmos.esa.int/gaia}), processed by the {\it Gaia}
Data Processing and Analysis Consortium (DPAC,
\url{https://www.cosmos.esa.int/web/gaia/dpac/consortium}). Funding for the DPAC
has been provided by national institutions, in particular the institutions
participating in the {\it Gaia} Multilateral Agreement.
This research has made use of the NASA/ IPAC Infrared Science Archive, which is 
operated by the Jet Propulsion Laboratory, California Institute of Technology, 
under contract with the National Aeronautics and Space Administration.
Based on observations collected at the European Organisation for Astronomical Research 
in the Southern Hemisphere under ESO programme(s) 090.C-0815 and 106.212F.001.
We used the VizieR catalogue access tool and the Simbad object data base at CDS to 
gather data. This work was supported by the Hungarian National Research, Development and 
Innovation Office grants OTKA K131508 and the \'Elvonal grant KKP-143986. KV 
is supported by the Bolyai J\'anos Research Scholarship of the Hungarian Academy of 
Sciences and by the Bolyai+ grant \'UNKP-22-5-ELTE-1093. SM is supported by a Royal Society
University Research Fellowship (URF-R1-221669). KYLSU acknowledges support from NASA 
ADAP programs (grant No. NNX17AF03G and 80NSSC20K1002).

\section*{Data Availability}
The {VISIR} data used in this paper are publicly available at ESO Archive (\url{http://archive.eso.org/eso/eso_archive_main.html}).



\bibliographystyle{mnras}
\bibliography{refs.bib} 





\appendix

\section{Membership of J071206 in the Crius\,228 moving group} \label{sec:appendix}
\begin{figure*}
    \centering
    \includegraphics[width=0.48\textwidth]{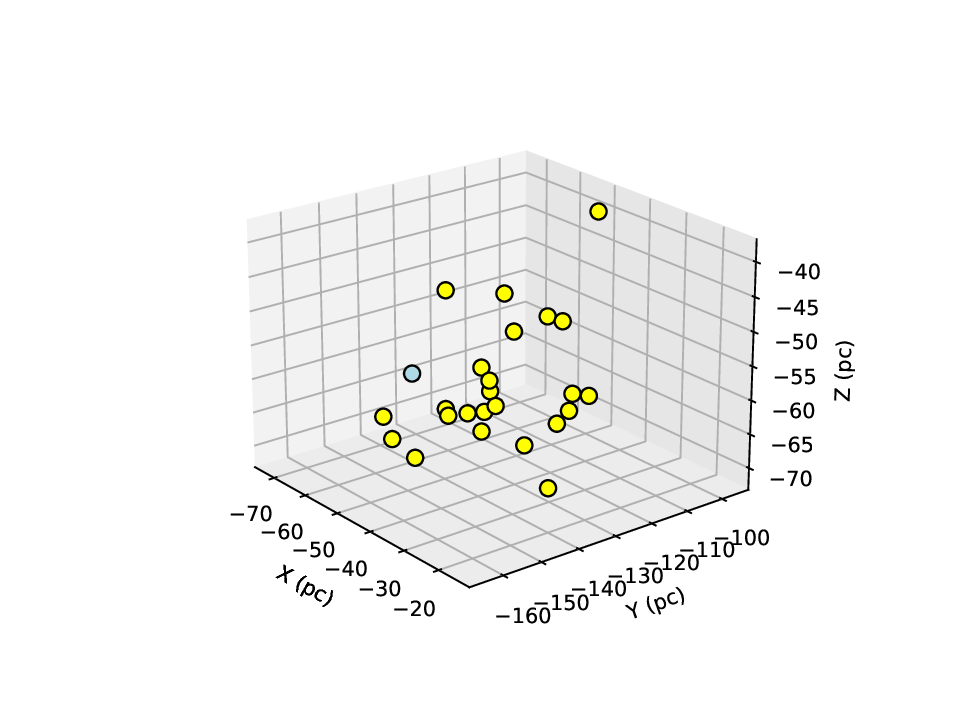}
    \includegraphics[width=0.48\textwidth]{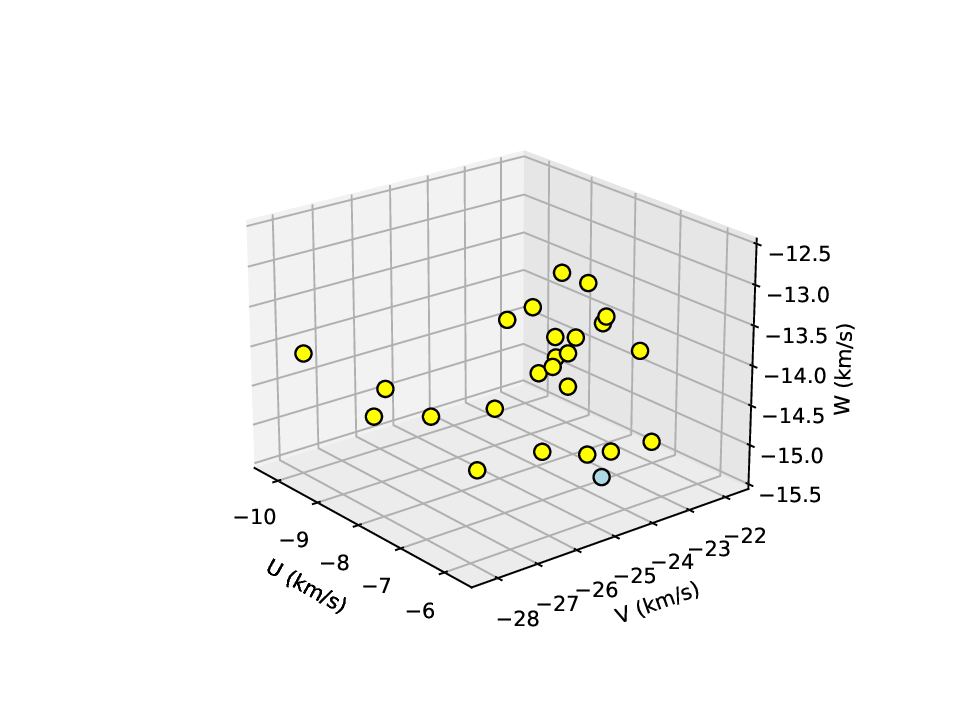}
    \includegraphics[width=0.44\textwidth]{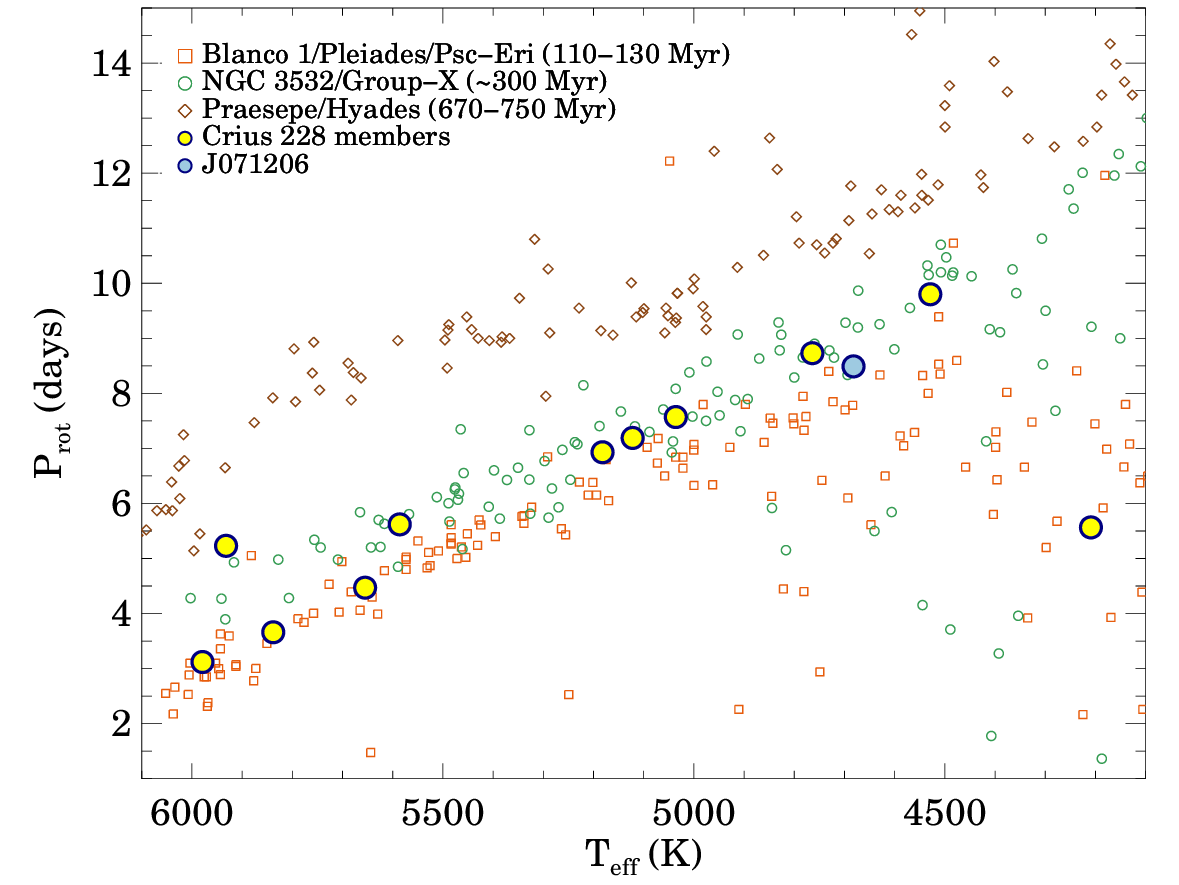}
    \includegraphics[width=0.44\textwidth]{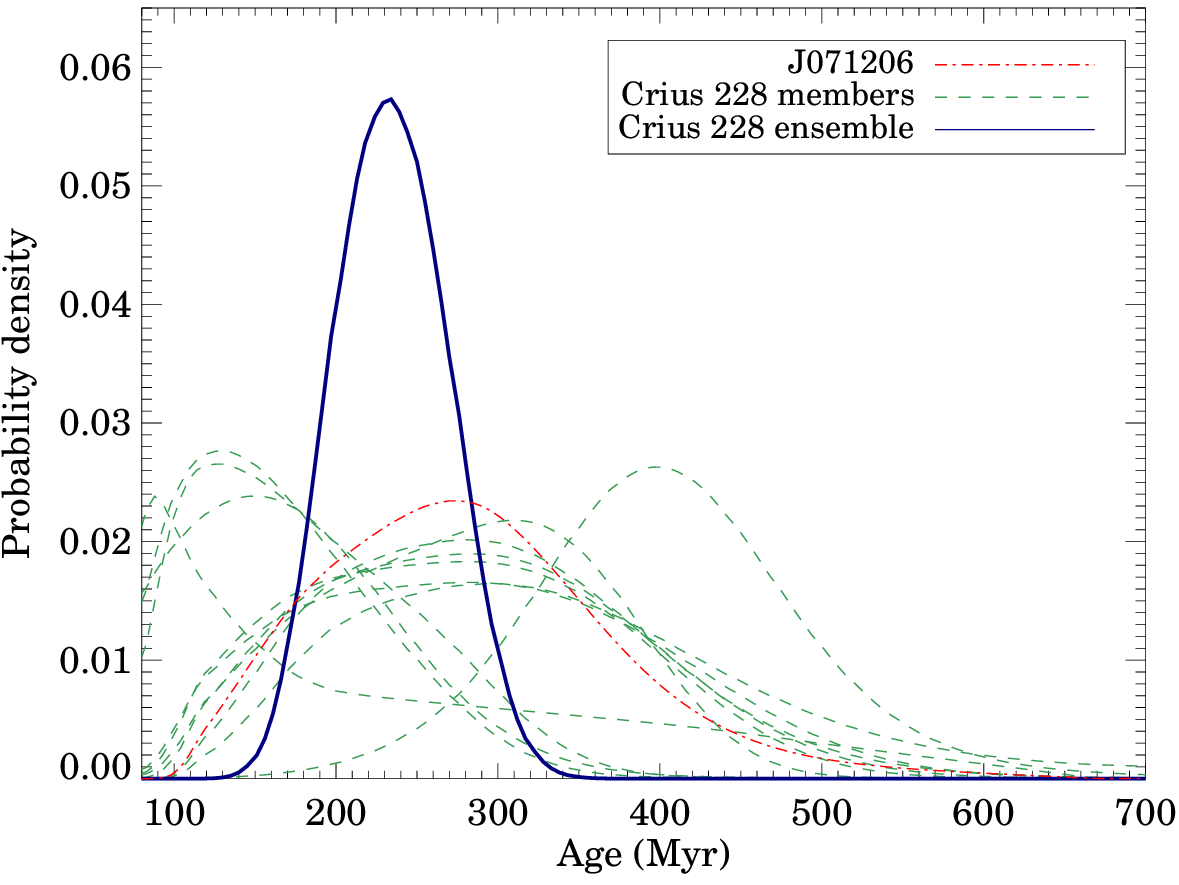}  
    \caption{Top panels: the X, Y, and Z galactic coordinates (left) and U, V, and W 
    galatic space velocity components (right) of Crius\,228 members \citep{moranta2022} with \texttt{ruwe} $<$1.2 (yellow circles)
    and J071206 (blue circle). Bottom panel, left: rotation periods of Crius\,228 members 
    and J071206 as a function of $T_\mathrm{eff}$. For comparison $T_\mathrm{eff}$-$P_\mathrm{rot}$ 
    distributions of 6 stellar groups are also displayed. The latter data were taken from \citet{bouma2023}.
    Bottom panel, right: age posteriors for individual stars from gyrochronology analysis performed 
    by the \textsc{gyro-interp} tool.
    }
    \label{fig:crius}
\end{figure*}

By analysing the Galactic space positions and velocities of Gaia Early Data Release 3 stars 
located within 200\,pc using the HDBSCAN unsupervised clustering algorithm, \citet{moranta2022} 
have identified 241 stellar clusters. As Figure~\ref{fig:crius} (top panels) demonstrates the location 
and motion of J071206 are similar to those of the Crius\,228 group members: 
it is located 36.2\,pc from the centre of the group and its space velocity 
components are U=$-$6.0$\pm$0.1\,km~s$^{-1}$, V=$-$24.6$\pm$0.1\,km~s$^{-1}$, and W=$-$15.1$\pm$0.2\,km~s$^{-1}$ 
(see Table~\ref{tab:props}), while the characteristic UVW velocities of the group 
are $-$7.5\,km~s$^{-1}$, $-24.7$\,km~s$^{-1}$, and $-$14.1\,km~s$^{-1}$, respectively.

To estimate the age of the Crius\,228, we used gyrochronology by employing the recent calibration 
proposed by \citet{bouma2023}. To use the latter method, we need to know the effective temperatures 
and rotation periods of the stars. The $T_\mathrm{eff}$ parameters were estimated from the 
dereddened Gaia $(B\mathrm{p}-R\mathrm{p})_0$ colour indices by adopting the conversion from \citet{curtis2020}.
Based on the Stilism 3D extinction map\footnote{\url{https://stilism.obspm.fr/}} \citep{lallement2014,capitanio2017}, we found that only one of the targets has non-negligible reddening. Since the applied gyrochronal model is validated for stars with $T_\mathrm{eff}$ between 3800\,K and 6200\,K \citep{bouma2023}, we removed all targets with temperatures out of this interval from the sample. 
Two additional stars, where the \texttt{ruwe} values of $>$1.2 indicate possible binarity, 
were also removed, as the presence of a nearby pair may affect their colour index and hence the 
$T_\mathrm{eff}$ estimate. To determine  the rotation period of the remaining fifteen targets, we 
analysed their TESS light curves following the method described in Sect.~\ref{sec:opticalvar}. These objects 
were measured in 2--13 TESS sectors, with data from 4 to 6 sectors for most. Finally, we found 11 
cases with well-defined periodic rotational modulation, the data of which are summarized in Table~\ref{tab:crius228}. Figure~\ref{fig:crius} (bottom left panel) shows the derived rotation 
periods as a function of $T_\mathrm{eff}$ for Crius\,228 members as well as for well dated 
clusters with ages between 120 and 670\,Myr for comparison. 
The majority of Crius\,228 stars constitute a quite narrow sequence in between the loci of the $\sim$120\,Myr 
old Pleiades/Blanco~I/Psc-Eri and the 300\,Myr old Group~X/NGC~3521 groups, suggesting an age of 200--250\,Myr.
To provide a more quantitative age estimate, we used the \textsc{gyro-interp} tool \citep{bouma2023} 
to derive age posteriors for each member using their $T_\mathrm{eff}$ and $P_\mathrm{rot}$ 
parameters, and then we determined the age posterior of the group by combining the individual ones 
(Figure~\ref{fig:crius}, bottom right panel). This yields an age estimate of 231$\pm$35\,Myr 
for Crius\,228. As shown in Fig.~\ref{fig:crius} (bottom panels) -- assuming that the rotation 
period obtained for J071206 is associated with the primary component and not to the putative 
secondary -- then this object fits very well with the rotation sequence of Crius\,228 members 
giving an additional indication of its membership.  
 
Using the excess photometric uncertainty in Gaia photometry as a proxy for variability, 
\citet{barber2023} found that the distribution of their variability parameter (the 90th percentile 
of the values) defined in this way for a stellar association correlates well with the age of 
the group and can thus be used for age diagostics. Though this method works best for groups with 
$>$100 members, by applying it for the relatively small Crius\,228 using the EVA tool\footnote{\url{https://github.com/madysonb/EVA}}, we obtained an age estimate of 
260$^{+68}_{-46}$\,Myr, which is good accordance with the gyrochrone age. 
\citet{moranta2022} proposed that the Crius\,228 possibly constitute a subset of the Theia\,301 association 
\citep{kounkel2019}. For the latter group, \citet{kounkel2020} derived an isochronal age of 195$^{+87}_{-60}$\,Myr, 
 which is also consistent with both of the above estimates. On the other hand, \citet{gagne2021} argued that 
 Theia\,301 may form a trailing tidal tail behind the Pleiades open cluster and has an age of $\sim$110\,Myr.

Further investigations are needed to see how Crius\,228 and Theia\,301 relate to each other and whether the 
latter is a single entity or a mixture of several clusters of different ages. The member list proposed by \citet{moranta2022} for Crius\,228 is based on the Gaia EDR3 catalogue. In the new DR3 database, radial 
velocity data are available for 
many more objects, the analysis of which may help to clarify the above question and possibly lead to the 
identification of additional Crius\,228 members thereby providing the opportunity to further refine the age 
of the group. Such a study, however, is out of the scope of the present work.

\begin{table}                                                                  
\begin{center} 
\caption{Rotation periods for Crius\,228 members \label{tab:crius228}}
\begin{tabular}{lccc}                                                     
\hline\hline
Gaia DR3 name & $T_\mathrm{eff}$ &  $P_\mathrm{rot}$ & TESS Sectors \\
\hline
2884927509794635392 & 4209 & 5.57$\pm$0.02 & 32,33 \\
2885816980343235968 & 5932 & 5.23$\pm$0.21 & 6,32,33 \\
2890261206342983680 & 5122 & 7.19$\pm$0.38 & 5,6,32,33 \\
5502604317431258624 & 5655 & 4.47$\pm$0.10 & 5,6,7,8,32,33,34,35\\
5504497161061675648 & 5838 & 3.66$\pm$0.09 & 6,7,8,33,34,35  \\
5507915130394590592 & 5979 & 3.12$\pm$0.14 & 6,7,8,33,34,35 \\
5551398959847312640 & 4529 & 9.80$\pm$0.60 & 6,7,8,33,34 \\
5553080555504933120 & 4764 & 8.73$\pm$0.28 & 5,7,28,32,33,34 \\
5555858785167602432 & 5586 & 5.62$\pm$0.15 & 6,7,33,34 \\
5566867125647181568 & 5036 & 7.57$\pm$0.12 & 5,7,32,33,34\\
5566896748038703488 & 5182 & 6.93$\pm$0.12 & 5,7,32,33,34\\
\hline
\\
\end{tabular}
\end{center}
\end{table}

\section{NEOWISE W1 and W2 light curves of recently identified EDDs} \label{sec:eddvar}
In this section, we present light curves showing how the {\sl WISE} $W1$ and $W2$ band disc fluxes of seven EDDs 
(J060917, J071206, J092521, J104416, J204315, J045959, and TYC\,8830) were changed on yearly timescale between 2013 and 2022.   

\begin{figure*}
    \centering
    \includegraphics[width=0.95\textwidth]{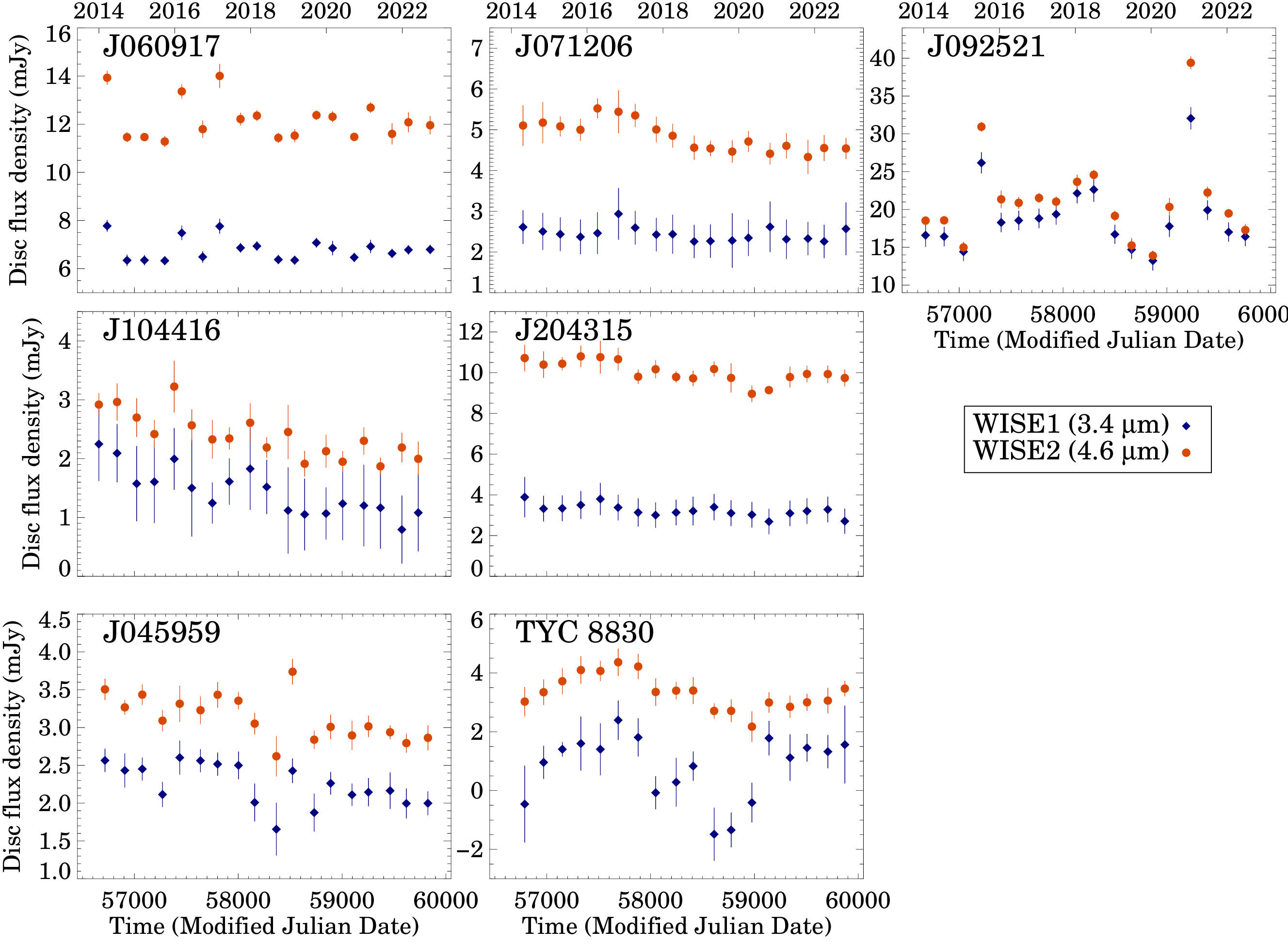}
    \caption{Disc flux data (the measured excess emissions) of the five EDDs analysed in Sect.~\ref{sec:newlydiscovered} 
    (top five panels) and two other recently identified EDDs (bottom two panels) in the 
    {\sl WISE} $W1$ and $W2$ bands from 2013 to 2022 during the NEOWISE Reactivation mission. The 
     displayed light curves have a cadence of $\sim$180\,d and their data points are
     derived by averaging the available good quality single exposure measurements in each 
     observing window (Sect.~\ref{sec:irvar}). 
    }
    \label{fig:irvar}
\end{figure*}

\section{The list of currently known EDDs}

\begin{table}                                                                  
\begin{center} 
\caption{List of known EDDs. References for discovery papers:  
1 - \citet{dewit2013}; 2 - \citet{gorlova2004}; 3 - \citet{gorlova2007}; 4 - \citet{higashio2022};
5 - \citet{mannings1998}; 6 - \citet{melis2012}; 7 - \citet{melis2021}; 8 - \citet{moor2021};
9 - \citet{oudmaijer1992}; 10 - \citet{rhee2008}; 11 - \citet{song2005}; 12 - \citet{tajiri2020};
13 - \citet{zuckerman2012}; 14 - This paper.
\label{tab:eddnames}}
\begin{tabular}{lccc}                                                     
\hline\hline
Short ID$^a$ & AllWISE ID & Simbad ID & Discovery \\
\hline
         RZ\,Psc &  J010942.07+275701.7 & V*\,RZ\,Psc     & 1 \\
      BD+20\,307 &  J015450.37+211822.2 & BD+20\,307      & 11 \\
       HD\,15407 &  J023050.76+553253.3 & HD\,15407       & 9 \\
       V488\,Per &  J032818.69+483947.9 & V*\,V488\,Per   & 13  \\
       HD\,23514 &  J034638.40+225510.7 & HD\,23514       & 10 \\
        J045959 &  J045959.96-084136.7  & TIC\,43488669   & 12 \\
       TYC\,4515 &  J050407.19+775857.1 & TYC\,4515-485-1 & 8 \\
       TYC 5940 &  J060513.59-191308.4 & TYC\,5940-1510-1  & 8 \\
       J060917 &  J060917.00-150808.5 &  -  &  14 \\
       TYC 8105 &  J061103.54-471129.2 & TYC\,8105-370-1   & 8 \\
       J071206 &  J071206.54-475242.3 &  -  &  14 \\
          P1121 &  J073542.68-145042.2  & Cl*\,NGC\,2422\,PMS\,1121 & 2 \\
            ID8 &  J080902.49-485817.2  & [GBR2007]\,ID\,8 & 3 \\
        J092521 &  J092521.90-673224.8 & TYC\,9196-2916-1   & 4 \\
	J104416 &  J104416.70-451613.9 & CD-44\,6765 & 14 \\
       TYC\,8241 &  J120902.21-512041.0 & TYC\,8241-2652-1 & 6 \\
       TYC 4946 &  J121334.13-053543.4 & TYC\,4946-1106-1   & 8 \\
      HD\,113766 &  J130635.75-460202.1 & HD\,113766   &  9 \\
      HD\,145263 &  J161055.09-253121.9 & HD\,145263   & 5 \\
      HD\,166191 &  J181030.32-233400.5 & HD\,166191   &  9 \\
       TYC\,4209 &  J181703.92+643354.9 & TYC\,4209-1322-1 & 8 \\
       J204315 &  J204315.23+104335.3 &  -  &  14 \\
       TYC\,8830 &  J230112.67-585821.9 & TYC\,8830-410-1  & 7 \\
       TYC\,4479 &  J235250.65+673037.2 & TYC\,4479-3-1    & 8 \\
\hline
\multicolumn{4}{l}{$^a$ Short identifiers of EDDs using throughout the paper.}\\
\\
\end{tabular}
\end{center}
\end{table}



\bsp	
\label{lastpage}
\end{document}